\documentclass{ar-1col_for_arXiv}
\usepackage[numbers]{natbib}
\usepackage{url}
\usepackage{amsmath} 
\usepackage[dvipdfmx,colorlinks=true,linkcolor=blue,citecolor=blue]{hyperref}
\jname{This article is posted with permission from the Annual Review of Condensed Matter Physics, Volume 14 \copyright \, by Annual Reviews, http://www.annualreviews.org.}

\begin{document}

\markboth{Kikkawa ${\cdot}$ Saitoh}{Spin Seebeck Effect}

\title{Spin Seebeck Effect: Sensitive Probe for Elementary Excitation, Spin Correlation, Transport, Magnetic Order, and Domains in Solids}
%

\author{Takashi Kikkawa$^1$ and Eiji Saitoh$^{1,2,3,4}$
\affil{$^1$Department of Applied Physics, The University of Tokyo, Tokyo 113-8656, Japan; email: t.kikkawa@ap.t.u-tokyo.ac.jp; ORCID: 0000-0002-7789-604X}
\affil{$^2$Institute for AI and Beyond, The University of Tokyo, Tokyo 113-8656, Japan}
\affil{$^3$WPI Advanced Institute for Materials Research, Tohoku University, Sendai 980-8577, Japan}
\affil{$^4$Advanced Science Research Center, Japan Atomic Energy Agency, Tokai 319-1195, Japan}}

\begin{abstract}
The spin Seebeck effect (SSE) refers to the generation of a spin current as a result of a temperature gradient in a magnetic material, which can be detected electrically via the inverse spin Hall effect in a metallic contact. 
Since the discovery of SSE in 2008, intensive studies on SSE have been conducted to elucidate its origin. 
SSEs appear in a wide range of magnetic materials including ferro-, ferri-, and antiferro-magnets and also paramagnets with classical or quantum spin fluctuation. 
SSE voltage reflects fundamental properties of a magnet, such as elementary excitation, static magnetic order, spin correlation, and spin transport. 
In this article, we review recent progress on SSEs in various systems, with particular emphasis on its emerging role as a probe of these magnetic properties in solids. 
We also briefly discuss the recently-discovered nuclear SSE. 
\end{abstract}

\maketitle

\tableofcontents

\section{INTRODUCTION}\label{sec:introduction}
\begin{figure}[bth]
\begin{center}
\includegraphics{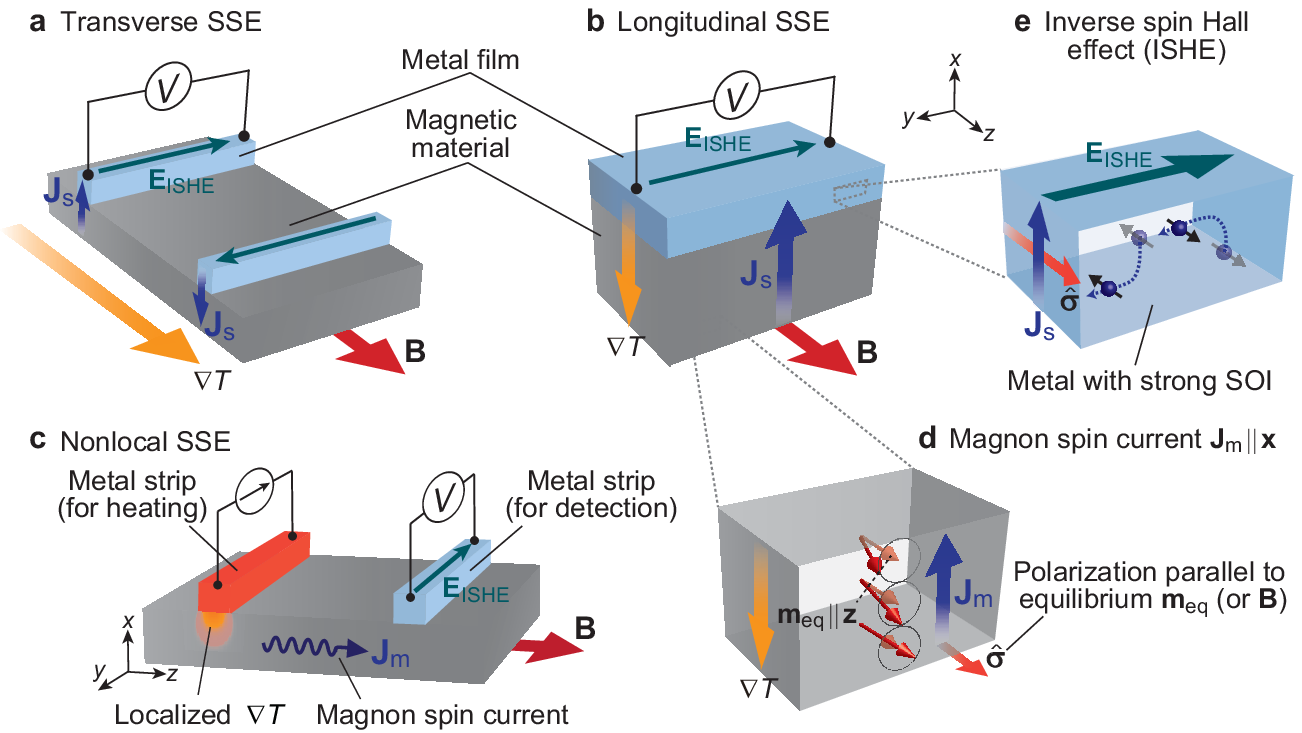}
\caption{Schematic illustrations of the (a) transverse SSE, (b) longitudinal SSE, (c) nonlocal SSE, (d) magnon spin current ${\bf J}_{\rm m}$, and (e) inverse spin Hall effect (ISHE). 
${\nabla} T$, ${\bf B}$, ${\bf E}_{\rm ISHE}$, ${\bf J}_{\rm s}$, and $\hat {\boldsymbol \sigma}$ denote the temperature gradient, external magnetic field (with magnitude $B$), electric field induced by the ISHE (with magnitude $E_{\rm ISHE}$), spatial direction of the thermally generated spin current, and spin polarization direction of the spin current (parallel to the equilibrium magnetization ${\bf m}_{\rm eq}$ of the magnetic layer), respectively. In the presence of the strong spin-orbit interaction (SOI), the spin current ${\bf J}_{\rm s}$ flowing in the metal layer is converted into a transverse voltage via the ISHE.  
}\label{fig:TSSE_LSSE_NLSSE_ISHE}
\end{center}
\end{figure}

The spin Seebeck effect (SSE) refers to the generation of a spin current, ${\bf J}_{\rm s}$, as a result of a temperature gradient, $\nabla T$, in a magnetic material with a metallic contact \cite{Uchida2014JPCM,Uchida2016ProcIEEE}. 
The effect was first discovered in permalloy in 2008 \cite{Uchida2008Nature} and later found in electrically insulating yttrium iron garnet (YIG; Y$_3$Fe$_5$O$_{12}$) \cite{Uchida2010NatMat} and ferromagnetic semiconductors (GaMnAs) \cite{Jaworski2010NatMat} in a configuration where a uniform $\nabla T$ is applied parallel to the magnetic film plane ({\bf Figure \ref{fig:TSSE_LSSE_NLSSE_ISHE}a}). 
In 2010, the most basic setup called the longitudinal configuration was demonstrated \cite{Uchida2010APL}, in which ${\nabla} T$ across a metal/magnet interface generates a spin current ${\bf J}_{\rm s}$ along ${\nabla} T$ ({\bf Figure \ref{fig:TSSE_LSSE_NLSSE_ISHE}b}). 
In 2015, the nonlocal SSE setup was introduced ({\bf Figure \ref{fig:TSSE_LSSE_NLSSE_ISHE}c}) \cite{Cornelissen2015NatPhys}, which can be applied to investigate spin-transport properties and further invigorated studies in spin caloritronics \cite{Bauer2012NatMat,Boona2014EnergyEnvironSci,Yu2017PhysLettA}.
\par
SSE voltages are generated in three steps: (i) the applied ${\nabla} T$ excites magnetization dynamics and thereby a magnon spin current ({\bf Figure \ref{fig:TSSE_LSSE_NLSSE_ISHE}d}) that (ii) at the interface to the metal becomes a conduction-electron spin current and (iii) is converted into a measurable voltage by the inverse spin Hall effect (ISHE) \cite{Spin-current-text} ({\bf Figure \ref{fig:TSSE_LSSE_NLSSE_ISHE}e}). 
Here, the ISHE is caused by the spin-orbit interaction, which bends electron orbits of up and down spins into opposite directions normal to their velocities. The resultant electric field ${\bf E}_{\rm ISHE}$ is given by
\begin{equation}\label{eq:ISHE}
{\bf E}_{\rm ISHE} \propto \theta _{\rm SH} {\bf J}_{\rm s} \times  \hat {\boldsymbol \sigma} ,
\end{equation}
where $\theta _{\rm SH}$, ${\bf J}_{\rm s}$, and $\hat {\boldsymbol \sigma}$ are the spin Hall angle, spatial direction of the injected spin current, and unit vector along the electron-spin polarization in the metallic layer (parallel to the equilibrium magnetization $\bf {m}_{\rm eq}$ of magnet), respectively ({\bf Figure \ref{fig:TSSE_LSSE_NLSSE_ISHE}e}) \cite{Uchida2016ProcIEEE}. Relatively high voltage is generated in heavy metals such as Pt, Ta, and W due to their large $\theta _{\rm SH}$ \cite{Spin-current-text}, allowing sensitive detection of SSEs. \par 
The longitudinal configuration  ({\bf Figure \ref{fig:TSSE_LSSE_NLSSE_ISHE}b}) has been mainly employed in recent studies owing to its simple and straightforward nature \cite{Uchida2014JPCM,Uchida2016ProcIEEE}, enabling systematic and quantitative investigations of SSEs in various magnetic insulators. 
Note that when a magnetic conductor is used in this configuration, anomalous Nernst effects may overlap with the longitudinal SSE (LSSE) signal \cite{Uchida2014JPCM,Uchida2016ProcIEEE,Kikkawa2013PRB,Bougiatioti2017PRL,De2020PRL}.
Materials used for LSSE measurements are listed in Figure 2 in the previous review article \cite{Uchida2016ProcIEEE}. 
Recent updates include
ferromagnetic EuO \cite{Mallick2019PRB}, 
two-dimensional (2D) ferromagnetic Cr$_2$Si$_2$Te$_6$, Cr$_2$Ge$_2$Te$_6$ \cite{Ito2019PRB,Gupta2020NanoLett}, 
ferrimagnetic garnet ferrites 
$R_3$Fe$_5$O$_{12}$ ($R$ = Eu, Tb, Dy, Tm) \cite{Cramer2017NanoLett,Avci2017PRB,Ortiz2021PRMater}, 
Y$_{3-x}R_x$Fe$_5$O$_{12}$ with $R$ being 14 rare-earth elements from La to Lu (except for Pm) \cite{Iwasaki2019SciRep},  %
Lu$_{2}$Bi$_{1}$Fe$_{4}$Ga$_{1}$O$_{12}$ \cite{Ramos2019NatCommun}, 
spinel ferrites
ZnFe$_2$O$_4$ \cite{Arboleda2016APL}, 
$\gamma$-Fe$_2$O$_3$ \cite{Jimenez-Cavero2017APLMater}, 
LiFe$_5$O$_8$ \cite{Shiomi2017SciRep},  
Ni$_{0.65}$Zn$_{0.35}$Al$_{0.8}$Fe$_{1.2}$O$_{4}$ \cite{Wang2018APL}, 
Mg$_{0.5-\delta}$Mn$_{0.5}$Fe$_{2}$O$_{4}$ \cite{Kosaki2021JPSJ}, 
Y-type hexagonal ferrites 
Ba$_{2}$Co$_{2}$Fe$_{12}$O$_{22}$, Ba$_{2}$Zn$_{2}$Fe$_{12}$O$_{22}$ \cite{Hirschner2017PRB}, 
orthorhombic ferrimagnetic $\varepsilon$-Fe$_2$O$_3$ \cite{Knizek2018JAP}, 
molecular-based ferrimagnetic Cr$^{\rm II}$[Cr$^{\rm III}$(CN)$_6$] \cite{Oh2021NatCommun},  
various antiferromagnetic (AF) insulators such as NiO \cite{Holanda2017APL,Ribeiro2019PRB,Gray2019PRX,Hoogeboom2020PRB}, %
FeF$_2$ \cite{Li2019PRL}, %
$\alpha$-Fe$_2$O$_3$ \cite{Yuan2020APL,Ross2021PRB}, %
MnCO$_3$ \cite{Kikkawa2021NatCommun}, %
$\alpha$-Cu$_2$V$_2$O$_7$ \cite{Shiomi2017PRB}, %
SrFeO$_3$ \cite{Hong2019APL}, %
SrMnO$_3$ \cite{Das2021APL}, %
DyFeO$_3$ \cite{Hoogeboom2021PRB}, %
and other intriguing materials including a chiral helimagnet Cu$_2$OSeO$_3$ with a skyrmion lattice phase \cite{Aqeel2016PRB,Akopyan2020PRB}  
and quantum magnets Sr$_2$CuO$_3$ \cite{Hirobe2017NatPhys,Hirobe2018JAP}, CuGeO$_3$ \cite{Chen2021NatCommun_CuGeO3}, Pb$_2$V$_3$O$_9$ \cite{Xing2022APL}, LiCuVO$_4$ \cite{Hirobe2019PRL}. 
In particular, the ferrimagnetic insulator YIG has been essential  \cite{Uchida2014JPCM,Uchida2016ProcIEEE}, as it exhibits the lowest magnetic damping, high Curie temperature ($T_{\rm C} \sim 560~\textrm{K}$), and high resistivity and also is a playground to reveal the role of magnon polarization in SSEs (see Section \ref{sec:magnon-polarization}). 
Experimental reports include 
temperature $T$ \cite{Uchida2014PRX,Kikkawa2015PRB,Jin2015PRB,Guo2016PRX,Iguchi2017PRB}, 
magnetic field $B$ \cite{Kikkawa2015PRB,Jin2015PRB,Guo2016PRX,Weiler2012PRL,Aqeel2014JAP,Uchida2015PRB,Kikkawa2016JPSJ,Kikkawa2016PRL,Kalappattil2017SciRep,Wu2020PRB}, 
length-scale (thickness) \cite{Kikkawa2015PRB,Jin2015PRB,Guo2016PRX,Kehlberger2015PRL,Miura2017PRMater,Prakash2018PRB,Daimon2020APEX}, 
structural \cite{Wu2018PRL,Nozue2018APL},
and time \cite{Roschewsky2014APL,Agrawal2014PRB,Kimling2017PRL,Bartell2017PRAppl,Hioki2017APEX,Seifert2018NatCommun,Jamison2019PRB} dependence measurements,  
investigation of the reciprocal effect \cite{Flipse2014PRL,Daimon2016NatCommun,Sola2019SciRep}, 
evaluation of a magnon temperature and chemical potential \cite{Agrawal2013PRL,Cornelissen2016PRB_chemical-potential,An2016PRL,Olsson2020PRX}, and so on. 
Some of the basic experimental results on LSSE for YIG (not focused here) are reviewed in References \cite{Uchida2014JPCM,Uchida2016ProcIEEE}, and we would like to ask interested readers to consult them. \par
%
%
SSE has been established as a universal phenomenon of magnetic materials, and recently received a lot of attention as a spectroscopic and tabletop tool to detect static and dynamical properties of magnets. 
In this article, we review recent progress on SSEs in various systems, with emphasis on its emerging role as a probe of elementary excitation, spin correlation, transport and associated scattering rates, static magnetic order and domains in solids. 
We start with a brief introduction to the mechanism of SSE in Section \ref{sec:theory}. 
From Sections \ref{sec:transport} to \ref{sec:qSSE}, we discuss SSE as a probe for dynamical and static properties of magnetic materials including paramagnets with classical or quantum spin fluctuation. 
In Section \ref{sec:NSSE}, we discuss the SSE driven by nuclear spins. 
Finally, we conclude this article and provide an outlook for further research opportunities. \par
%
%
\section{THEORETICAL ESSENCE OF SPIN SEEBECK EFFECT}\label{sec:theory}
\begin{figure}[tbh]
\begin{center}
\includegraphics{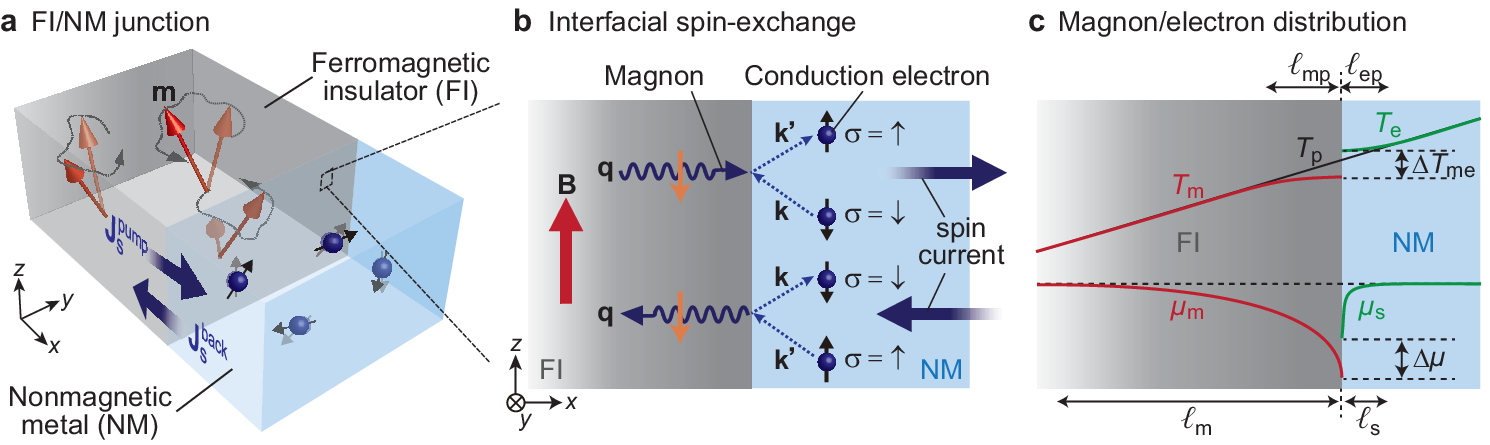}
\caption{
(a) FI/NM hybrid structure and spin-pumping current ${\bf J}_{\rm s}^{\rm pump}$ (backflow spin current ${\bf J}_{\rm s}^{\rm back}$) driven by thermal activation of the magnetization ${\bf m}$ in FI (electron spins in NM). 
At thermal equilibrium, ${\bf J}_{\rm s}^{\rm pump}$ and ${\bf J}_{\rm s}^{\rm back}$ compensate \cite{Xiao2010PRB}. %
(b) Interfacial spin-exchange coupling between FI and NM.  
An electron in NM undergoes a spin-flip inelastic scattering at the interface, creating or annihilating a magnon in FI, which causes an interfacial spin current, $J_{\rm s}^{\rm int}$  \cite{Bender2012PRL}. 
(c) A schematic profile of the magnon temperature $T_{\rm m}$, electron temperature $T_{\rm e}$, magnon chemical potential $\mu_{\rm m}$, and electron spin accumulation $\mu_{\rm s}$, assuming a constant phonon temperature gradient ${\bf \nabla}T_{\rm p}$ and disregarding the interfacial heat (Kapitza) resistance. 
$T_{\rm m}$ ($T_{\rm e}$) and $\mu_{\rm m}$ ($\mu_{\rm s}$) relax on length scales $\ell_{\rm mp}$ ($\ell_{\rm ep}$) and $\ell_{\rm m}$ ($\ell_{\rm s}$), respectively, which are governed by heat and magnon (spin) diffusion equations \cite{Cornelissen2016PRB_chemical-potential}. 
For YIG/Pt at room temperature, these length scales are typically $\ell_{\rm mp}^{\rm YIG} \sim$ of the order of a nm, $\ell_{\rm m}^{\rm YIG} \sim 10~\mu \textrm{m}$,  $\ell_{\rm ep}^{\rm Pt} \sim 5~\textrm{nm}$, and $\ell_{\rm s}^{\rm Pt} \sim 2~\textrm{nm}$ 
\cite{Cornelissen2016PRB_chemical-potential}.
}\label{fig:SSE_mechanism}
\end{center}
\end{figure}

A theoretical formulation of the magnon-driven SSE was developed by Xiao et al. in 2010 \cite{Xiao2010PRB}. 
Let us first consider a ferromagnetic insulator (FI) at thermal equilibrium with an attached nonmagnetic metal (NM) (see {\bf Figure \ref{fig:SSE_mechanism}a}). 
When the FI is thermally excited, the dynamics of magnetization ${\bf M}(t)$ (with unit vector ${\bf m}(t)$) injects a dc spin current into the NM due to spin pumping \cite{Spin-current-text}, 
\begin{equation}\label{equ:interfacial-SSE_Xiao1}
\langle {\bf J}_{\rm s}^{\rm pump} \rangle _z
= \frac{\hbar g_r^{\uparrow \downarrow}}{4 \pi} \langle {\bf m} \times {\dot {\bf m}} \rangle _z ,
\end{equation} 
which is proportional to the real part of the interfacial spin-mixing conductance $g_r^{\uparrow \downarrow}$ and equal-time and space spin correlation functions or transverse dynamical susceptibility of the FI at the interface \cite{Xiao2010PRB,Barker2016PRL}. 
At finite temperatures, however, thermal (Johnson--Nyquist) noise in the NM generates a backflow spin current $\langle {\bf J}_{\rm s}^{\rm back} \rangle _z $, which compensates for the pumped spin current $\langle {\bf J}_{\rm s}^{\rm pump} \rangle _z$ on average at thermal equilibrium, satisfying the second law of thermodynamics \cite{Bauer2012NatMat}
The fluctuation-dissipation theorem shows that the thermal spin current $\langle {\bf J}_{\rm s}^{\rm pump} \rangle _z$ from FI to NM ($\langle {\bf J}_{\rm s}^{\rm back} \rangle _z $ from NM to FI) is in proportion to the effective magnon (electron) temperature 
$T_{\rm m}$ ($T_{\rm e}$) in the FI (NM), leading to the net dc component \cite{Uchida2010NatMat,Xiao2010PRB}
\begin{equation}\label{equ:interfacial-SSE_Xiao2}
J_{\rm s}^{\rm int}
= \langle {\bf J}_{\rm s}^{\rm pump} \rangle _z - \langle {\bf J}_{\rm s}^{\rm back} \rangle _z
= 2 \alpha^{(1)} k_{\rm B} (T_{\rm m} - T_{\rm e}) , 
\end{equation} 
where $\alpha^{(1)} (\propto g_r^{\uparrow \downarrow})$ is the enhanced damping due to spin pumping. This equation shows that a finite interfacial spin current $J_{\rm s}^{\rm int}$ is generated when an effective magnon-electron temperature difference is induced by an external temperature gradient. Adachi et al. also derived a similar expression by linear response theory \cite{Adachi2011PRB,Adachi2013RepProgPhys}. \par
A perturbative treatment of the exchange coupling at the interface is also applied to describe the interfacial SSE \cite{Bender2012PRL}. %
Here, conduction electrons in the NM are coupled to the magnetic moments ${\bf m}$ in the FI via exchange interaction, of which the Hamiltonian is written in terms of the creation (annihilation) operators $a_{{\mathbf{q}}}^{\dagger }$ ($a_{{\mathbf{q}}}$) for Holstein-Primakoff magnons and $c_{{\bf k},\sigma }^{\dag}$ ($c_{{\bf k},\sigma }$) for conduction electrons \cite{Bender2012PRL}:
\begin{equation}\label{eq:SSE-interfacial-Hamiltonian}
{\mathcal H}_{\rm int} 
= \sum_{{\mathbf{q}} {\mathbf{k}} {\mathbf{k}}^{\prime}} V_{{\mathbf{q}} {\mathbf{k}} {\mathbf{k}}^{\prime}} a_{{\mathbf{q}}} c_{{\bf k}^{\prime},\uparrow}^{\dag} c_{{\bf k},\downarrow } \; + \; {\rm H.c.} , 
\end{equation}
where $\sigma = \; \uparrow$ ($\downarrow$) denotes the electron spin pointing along the $+z$ ($-z$) direction (see {\bf Figure \ref{fig:SSE_mechanism}b}). 
Calculation of the spin-flip rate at the interface based on Fermi's Golden Rule leads to the interfacial spin current per interfacial area $A$ ($J_{\rm s}^{\rm int} = A^{-1}\langle d s_z/dt \rangle$) %
\begin{equation}\label{eq:SSE-interfacial-spin-current}
J_{\rm s}^{\rm int} = - \frac{g_r^{\uparrow \downarrow}}{\pi s}\int^{\infty}_{\epsilon _0}d\epsilon \: {\mathcal D}(\epsilon) \:  (\epsilon - \mu_{\rm s})\:  
\left[ \, 
f_{\rm BE}(\epsilon, \mu_{\rm m}, T_{\rm m})
-
f_{\rm BE}(\epsilon, \mu_{\rm s}, T_{\rm e}) 
\,  \right] , 
\end{equation} %
where $s=S/a_0^3$ is the equilibrium spin density of the FI with $S$ being the total spin in a unit cell with volume $a_0^3$. 
Here, 
\begin{equation}\label{equ:Bose-Einstein}
f_{\rm BE}(\epsilon, \mu_{\rm m(s)}, T_{\rm m(e)})
=
\left[ \exp \left( \frac{\epsilon - \mu_{\rm m(s)}}{k_{\rm B}T_{\rm m(e)}} \right) - 1\right]^{-1}
\end{equation}
is the Bose--Einstein function describing the distribution for magnons (electron spin accumulation) in the FI (NM) that is parameterized by the magnon temperature $T_{\rm m}$ and chemical potential $\mu_{\rm m}$ (electron temperature $T_{\rm e}$ and spin accumulation  $\mu_{\rm s}$) 
({\bf Figure \ref{fig:SSE_mechanism}c}) \cite{Cornelissen2016PRB_chemical-potential,Bender2012PRL,Duine-text}. 
${\mathcal D}(\epsilon)$ is the density of states of magnons, and given as ${\mathcal D}(\epsilon)=[4\pi^2 (\hbar D_{\rm ex})^{\frac{3}{2}}]^{-1}\sqrt{\epsilon - \hbar \gamma B}$
for a parabolic dispersion $\epsilon = \hbar \omega_{\bf k} = \hbar D_{\rm ex}{\bf k}^2+\hbar \gamma B$ ($D_{\rm ex}$: exchange stiffness, $\gamma$: electron gyromagnetic ratio). 
The first (second) term $\propto \int d\epsilon {\mathcal D}(\epsilon) (\epsilon - \mu_{\rm s})\: f_{\rm BE}(\epsilon, \mu_{\rm m}, T_{\rm m})$ $\left( \propto \int d\epsilon {\mathcal D}(\epsilon) (\epsilon - \mu_{\rm s})\: f_{\rm BE}(\epsilon, \mu_{\rm s}, T_{\rm e}) \right)$ in Equation \ref{eq:SSE-interfacial-spin-current} represents the spin-pumping (backflow) contribution from the FI (NM) due to the thermal fluctuation of magnetization (itinerant electron spins) \cite{Duine-text}, as schematically shown in {\bf Figure \ref{fig:SSE_mechanism}a}. 
$\mu_{\rm m}$ in Equation \ref{equ:Bose-Einstein} parametrizes magnon accumulation and/or depletion states that cannot be expressed solely by a local $T_{\rm m}$ \cite{Cornelissen2016PRB_chemical-potential}. 
Cornelissen et al. \cite{Cornelissen2016PRB_chemical-potential} showed that, under the reasonable assumptions that magnons thermalize well with phonons and also that magnon-number conserving scatterings are stronger than magnon-number {\it non}-conserving scatterings, or magnon decay scatterings characterized by Gilbert damping $\alpha$, the introduction of $\mu_{\rm m}$ in magnon distribution may give a better explanation for magnon transport phenomena in terms of the length scale (see {\bf Figure \ref{fig:SSE_mechanism}c} and Section \ref{sec:transport}). 
In linear response, Equation \ref{eq:SSE-interfacial-spin-current} yields
\begin{equation}\label{eq:SSE-interfacial-spin-current-linear-response}
J_{\rm s}^{\rm int} =
\frac{\sigma_{\rm s}^{\rm int}}{\hbar \Lambda} (\mu_{\rm s}-\mu_{\rm m}) 
+
\frac{L_{\rm SSE}^{\rm int}}{\Lambda} (T_{\rm e}-T_{\rm m}) , 
\end{equation}
where $\sigma_{\rm s}^{\rm int} = 3 \zeta(3/2)\hbar g_r^{\uparrow \downarrow}/2\pi s \Lambda ^2$ is the interfacial spin conductivity and  $L_{\rm SSE}^{\rm int} = 15 \zeta(5/2)k_{\rm B} g_r^{\uparrow \downarrow}/4\pi s \Lambda ^2$ is the interfacial spin Seebeck coefficient with $\Lambda =\sqrt{4\pi \hbar D_{\rm ex}/k_{\rm B}T_{\rm m}}$ being the thermal de Broglie wavelength \cite{Cornelissen2016PRB_chemical-potential,Duine-text}. The above expression agrees with that derived from the stochastic model by Xiao et al. \cite{Xiao2010PRB}. \par 
%
%
%
Bulk magnon transport in FI also affects the interfacial magnon distribution (Equation \ref{equ:Bose-Einstein}) as it causes a non-equilibrium magnon accumulation and/or depletion at the interface with NM. A diffusive transport picture should be valid when the FI system size is larger than the magnon mean-free path and magnon thermal wavelength \cite{Cornelissen2016PRB_chemical-potential}. In 2012 and 2014, Zhang and Zhang \cite{Zhang-Zhang2012PRB} and Rezende et al. \cite{Rezende2014PRB,Rezende_text} pioneered bulk SSE theories based on diffusion equations of magnons. Subsequently, Cornelissen et al. \cite{Cornelissen2016PRB_chemical-potential} developed a diffusive magnon spin and heat transport theory based on linearized Boltzmann equations and obtained the magnon spin current ${\bf J}_{\rm m}$ and heat current ${\bf J}_{\rm Q,m}$ densities in FI as
\begin{equation}\label{eq:magnonic-currents}
\left(
\begin{array}
[c]{c}%
\frac{2e}{\hbar} {\bf J}_{\rm m}\\
{\bf J}_{\rm Q,m}
\end{array}
\right)  = - \left(
\begin{array}
[c]{cc}%
\sigma_{\rm m} & L/T \\
\hbar L/2e & \kappa_{\rm m}%
\end{array}
\right)  \left(
\begin{array}
[c]{c}%
{\bf \nabla} \mu_{\rm m}\\
{\bf \nabla} T_{\rm m}
\end{array}
\right) ,
\end{equation}
where $\sigma_{\rm m}$ is the magnon spin conductivity, $L$ is the bulk spin Seebeck coefficient, and $\kappa_{\rm m}$ is the magnonic heat conductivity \cite{Cornelissen2016PRB_chemical-potential}. 
Here, the distributions of $T_{\rm m}$ and $\mu_{\rm m}$ are governed by diffusion equations with characteristic relaxation lengths (see {\bf Figure \ref{fig:SSE_mechanism}c}), which are linked with the electron spin diffusion equation in NM by the interfacial boundary condition described as Equation \ref{eq:SSE-interfacial-spin-current} \cite{Cornelissen2016PRB_chemical-potential}. \par
\section{SSE AS PROBE OF MAGNON TRANSPORT}\label{sec:transport}
%
%
The magnon transport relevant to SSEs was first investigated through the thickness $t_{\rm FI}$ dependence of magnetic layer in the longitudinal configuration ({\bf Figure \ref{fig:TSSE_LSSE_NLSSE_ISHE}b}), which clarified the role of bulk magnon transport in SSEs. 
Length-scale evaluation based on a phenomenological model $\left( E_{\rm ISHE} \propto 1 - {\rm exp}(-t_{\rm FI}/ \xi) \right)$ yields $\xi \sim 1 ~\mu \textrm{m}$ for YIG/Pt systems at room temperature \cite{Kikkawa2015PRB,Guo2016PRX} as well as its $T$-scaling of $\xi \propto T^{-1}$ \cite{Guo2016PRX}.  
In 2018, Prakash et al. \cite{Prakash2018PRB} pointed out the existence of two distinct length scales, a magnon spin-diffusion length ($\ell_{\rm m} \sim 10~\mu\textrm{m}$) and a magnon energy relaxation length ($\ell_{\rm u} \sim 250~\textrm{mm}$), through the analysis of their $t_{\rm FI}$ dependent data for YIG/Pt systems by a magnon diffusion model. 
In 2020, based on optothermal imaging, Daimon et al. \cite{Daimon2020APEX} reported the spatially resolved LSSE voltage in a single YIG/Pt system with a $t_{\rm FI}$ gradient and found that the $t_{\rm FI}$ dependence is identical to that for the (reciprocal) spin Peltier effect (SPE), in which $\ell_{\rm m}$ is evaluated as $3.9~ \mu\textrm{m}$.  
Systematic $t_{\rm FI}$-dependent LSSE studies are reported also for Fe$_3$O$_4$ \cite{Anadon2016APL,Venkat2020PRMater}, whose magnon diffusion length is evaluated to be several tens nm. 
Ramos et al. \cite{Ramos2015PRB} found the enhanced LSSE in multilayered [Fe$_3$O$_4$/Pt]$_n$ by increasing the stacking number $n$ and explained the result by introducing a length associated with the multilayered structure. \par
\begin{figure}[tbh]
\begin{center}
\includegraphics{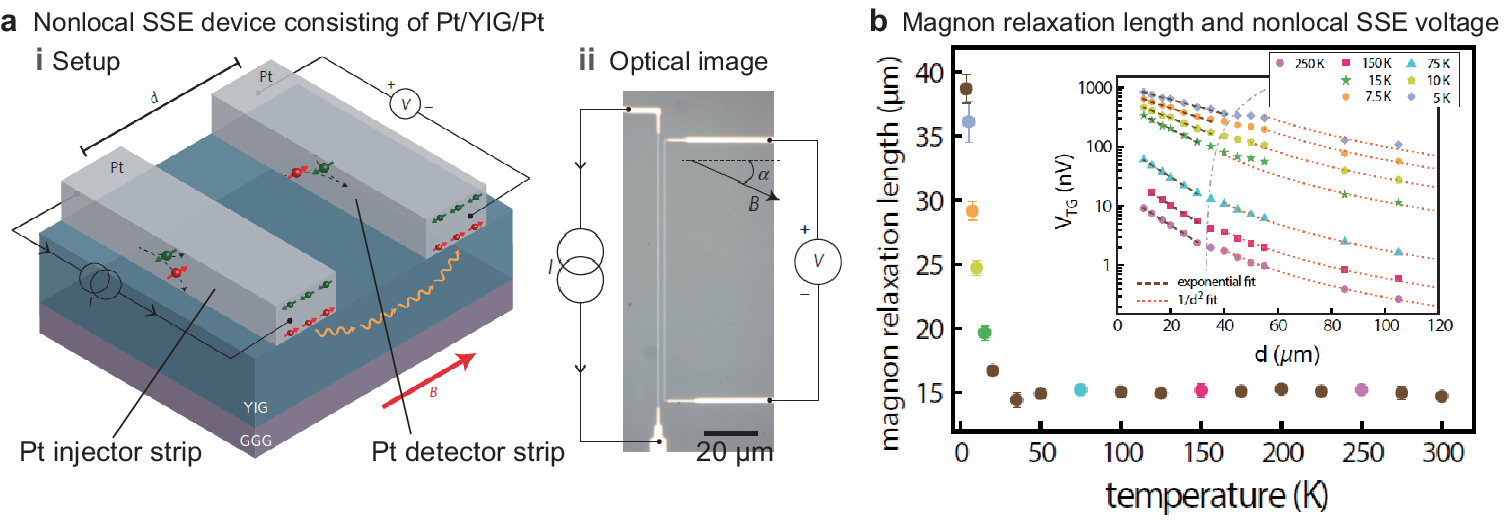}
\caption{
(a) ({\it i}) Setup for detecting the nonlocal SSE and ({\it ii}) optical microscope image for a nonlocal SSE device comprising two Pt injector and detector strips fabricated on an YIG film. 
Figure reproduced with permission from Reference \cite{Cornelissen2015NatPhys}. Copyright \copyright \, 2015 Springer Nature. 
(b) Nonlocal SSE voltage $V_{\rm TG}$ as a function of $d$ at various temperatures $T$ for Pt/YIG/Pt devices (inset) and the magnon relaxation length $\ell_{\rm m}$ extracted from exponential fits (main). 
Figure reproduced with permission from Reference \cite{Shan2017PRB}. Copyright \copyright \, 2017 American Physical Society. 
}\label{fig:Nonlocal-SSE}
\end{center}
\end{figure}
The nonlocal geometry ({\bf Figure \ref{fig:TSSE_LSSE_NLSSE_ISHE}c}) provides an alternative approach to address the (lateral) transport of thermally excited magnons \cite{Cornelissen2015NatPhys,Cornelissen2016PRB_chemical-potential}.  
{\bf Figure \ref{fig:Nonlocal-SSE}a} displays a sketch and optical microscope image of a typical nonlocal SSE device, consisting of two NM (= Pt) strips formed on FI (= YIG) that are electrically isolated from each other by a center-to-center distance $d$ \cite{Cornelissen2015NatPhys}. 
By applying a charge current to one of the NM (injector) strip, due to its local Joule heating, magnons are thermally excited in the FI layer beneath the strip and diffuse toward the other NM (detector) strip.  
When a part of them successfully reaches the detector strip, a spin current is injected into the strip and is subsequently converted into an electric voltage by the ISHE. 
This implies that the distance over which the magnons diffuse is well defined as the separation distance $d$ (typically, $d= 500~\textrm{nm} \sim 100~\mu \textrm{m}$), allowing us to extract the magnon spin relaxation length $\ell_{\rm m}$ via systematic nonlocal SSE measurements as a function of $d$ \cite{Cornelissen2016PRB-T-dep}. 
A laser heating method is also used to excite magnons, and their diffusion is nonlocally detected in a similar way as described here \cite{Giles2015PRB,Giles2017PRB}. 
\par
The inset to {\bf Figure \ref{fig:Nonlocal-SSE}b} shows the nonlocal SSE voltage $V_{\rm TG}$ versus $d$ for Pt/YIG/Pt nanofabricated devices 
\cite{Shan2017PRB}.  
$V_{\rm TG}$ monotonically decreases with increasing  $d$, which corresponds to the decay of the magnon spin signal governed by the magnon spin relaxation length $\ell_{\rm m}$ \cite{Cornelissen2016PRB-T-dep}. 
The extracted $\ell _{\rm m}$ values increase with decreasing $T$, and reach $\sim 40~\mu \textrm{m}$ at $3.5~\textrm{K}$ ({\bf Figure \ref{fig:Nonlocal-SSE}b}) \cite{Shan2017PRB}. The $\ell_{\rm m}$ estimation may be affected by the locally distributed $\nabla T$, which is discussed in References \cite{Shan2017PRB,An2021PRB}. 
So far, nonlocal SSEs are reported in several magnets such as 
ferrimagnetic YIG \cite{Cornelissen2016PRB-T-dep,Shan2017PRB,An2021PRB,Giles2015PRB,Giles2017PRB,Cornelissen2016PRB_H-dep,Shan2016PRB,Zhou2017APL,Ganzhorn2017AIPadv,Cornelissen2017PRB,Oyanagi2020AIPAdv,Gomez-Perez2020PRB}, 
Gd$_3$Fe$_5$O$_{12}$ \cite{Ganzhorn2017arXiv}, 
Tm$_3$Fe$_5$O$_{12}$ \cite{Avci2020PRL}, 
NiFe$_2$O$_4$ \cite{Shan2018APL}, 
MgAl$_{0.5}$Fe$_{1.5}$O$_4$ \cite{Li2022NanoLett}, 
antiferromagnetic Cr$_2$O$_3$ \cite{Yuan2018SciAdv,Muduli2021AIPAdv}, 
$\alpha$-Fe$_2$O$_3$ \cite{Ross2021PRB,Lebrun2018Nature}, 
NiO \cite{Hoogeboom2020PRB}, 
YFeO$_3$ \cite{Das2021arXiv}, 
BiFeO$_3$ \cite{Parsonnet2022arXiv}, 
and 2D layered magnets MnPS$_3$ \cite{Xing2019PRX,Chen2021NatCommun_MnPS3} and CrBr$_3$ \cite{Liu2020PRB}. \par
We note that the applied charge current in the injector Pt strip also induces a spin accumulation via the SHE \cite{Spin-current-text} at the interface to YIG, which excites nonequilibrium magnons through the interfacial spin-flip scattering ({\bf Figure \ref{fig:SSE_mechanism}b}). 
The electrically-driven magnon flows in YIG, and is finally converted into an ISHE voltage at the detector strip.   
This all-electrical magnon transport has also been investigated intensively, and recent progress is summarized in Reference \cite{Althammer2021PhysStatSolid}. \par  
%
%
%
\section{SSE AS PROBE OF MAGNETIC ORDER AND DOMAINS}\label{sec:magnetic-order}
SSE enables electric readout of the magnetic orientation near the magnet/metal interface. 
This feature is realized by the characteristic of ISHE electric field induced by SSEs: ${\bf E}_{\rm ISHE} \propto {\bf J}_{\rm s} \times  \hat {\boldsymbol \sigma}$ (Equation \ref{eq:ISHE}) \cite{Aqeel2014JAP,Uchida2015PRB}.  
In particular, for LSSEs, the spatial directions of ${\bf J}_{\rm s}$ and $\hat {\boldsymbol \sigma}$ in NM are, respectively, along the applied $\nabla T$ and equilibrium magnetization ${\bf m}_{\rm eq}$ near the interface, meaning that the output field satisfies the relationship ${\bf E}_{\rm ISHE} \propto {\nabla T} \times {\bf m}_{\rm eq}$.
The LSSE voltage can therefore extract the projection of the (surface) magnetization on the direction along the magnet/metal interface, allowing to reveal the in-plane magnetic orientation (within $\ell_{\rm m}$) and also magnetic anisotropies \cite{Aqeel2014JAP,Uchida2015PRB,Kalappattil2017SciRep,Wu2020PRB,Li2019APL,Chanda2022AdvFunctMater}. \par 
\begin{figure}[tb]
\begin{center}
\includegraphics{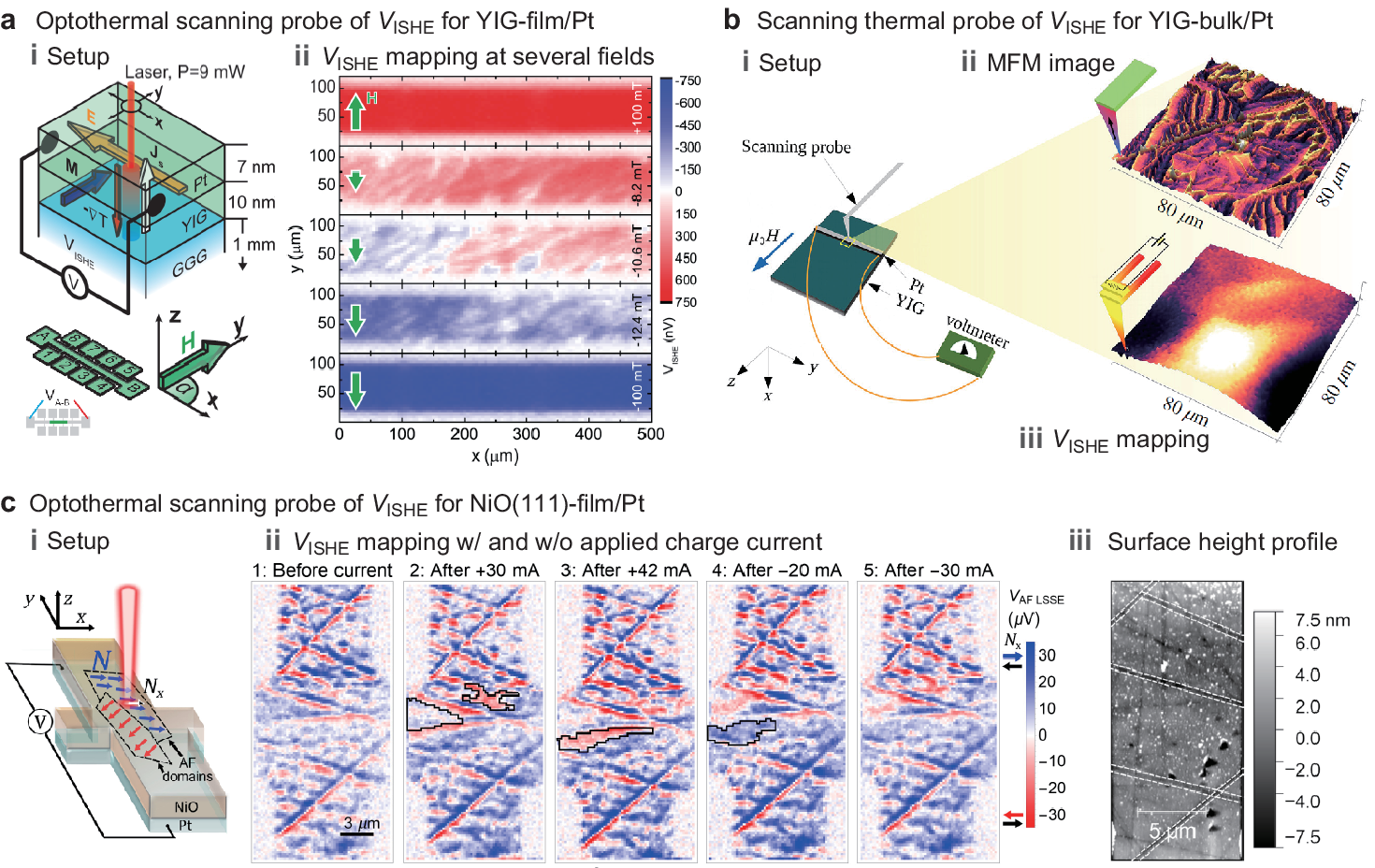}
\caption{
(a) ({\it i}) Setup for detecting the LSSE in an YIG-film/Pt system with a scannable laser beam.  
({\it ii}) ISHE voltage $V_{\rm ISHE}$ as a function of the laser-spot position $(x,y)$ at several fields. 
Figure reproduced with permission from Reference \cite{Weiler2012PRL}. Copyright \copyright \, 2012 American Physical Society. 
(b) ({\it i}) Setup for detecting the LSSE in an YIG-bulk/Pt system with a scanning thermal probe. 
 ({\it i}, {\it ii}) Comparison between the ({\it i}) MFM image and ({\it ii}) $V_{\rm ISHE}$ mapping data. 
Figure reproduced with permission from Reference \cite{Sola2020PRAppl}. Copyright \copyright \, 2020 American Physical Society. 
(c) ({\it i}) Setup for detecting the LSSE in a NiO(111)-film/Pt system with a scannable laser beam.  
({\it ii}) $V_{\rm ISHE}$ mapping with and without the application of charge current to the Pt film. Blue (red) area represents interfacial spins pointing right (left), some of which are switched with antidamping spin torque \cite{Gray2019PRX}.
Sharp straight lines are attributed to the signal originating from polishing scratches in the MgO (111) substrate that are also visible in the atomic force microscope image shown in ({\it iii}).
Figure reproduced with permission from Reference \cite{Gray2019PRX}. Copyright \copyright \, 2019 Authors, licensed under a Creative Commons Attribution (CC BY) license. 
}\label{fig:Magnetic-ordering_imaging}
\end{center}
\end{figure}
By utilizing the above feature, magnetic domain structures were detected through LSSE measurements \cite{Gray2019PRX,Weiler2012PRL,Bartell2017PRAppl,Sola2020PRAppl}.  
Using a scannable laser beam to create a local $\nabla T$ on an YIG-film/Pt bilayer, Weiler et al. \cite{Weiler2012PRL} demonstrated a spatial mapping of in-plane magnetic structure of the YIG ({\bf Figure \ref{fig:Magnetic-ordering_imaging}a}). 
Bartell et al. \cite{Bartell2017PRAppl} extended the approach into the time domain and addressed spatiotemporal thermal evolution of the LSSE in an YIG-film/Pt bilayer with sub-100 ps resolution.  
Sola et al. \cite{Sola2020PRAppl}, instead, used a scanning thermal probe consisting of a micropatterned cantilever and heated up the tip that is in contact with an YIG-bulk/Pt sample to create a local $\nabla T$. 
By scanning the cantilever tip, they observed a spatially resolved voltage response that depends on the magnetic domain distribution confirmed by magnetic force microscopy (MFM)  ({\bf Figure \ref{fig:Magnetic-ordering_imaging}b}). 
In 2019, Gray et al. \cite{Gray2019PRX} applied the optothermal imaging method for NiO-film/Pt systems, and obtained AF domain distributions at zero field and room temperature ({\bf Figure \ref{fig:Magnetic-ordering_imaging}c}), where the LSSE signal was attributed to the N\'eel order contribution. 
They also successfully visualized the domain rotation and domain wall motion due to the antidamping spin torque induced by the SHE in Pt ({\bf Figure \ref{fig:Magnetic-ordering_imaging}c}, {\it subpanel ii}). 
The technique may provide a simple and versatile way to characterize AF domains and to understand the spin-torque switching in antiferromagnets (AFMs).  \par 
We would like to note that thermal imaging of the reciprocal process of SSE (i.e., the SPE \cite{Daimon2020APEX,Flipse2014PRL,Daimon2016NatCommun,Sola2019SciRep,Yahiro2020PRB}) can also sensitively detect the magnetic orientation and domains. In 2018, Yagmur et al. \cite{Yagmur2018JPhysD} identified the magnetization distributions for ferrimagnetic GdIG across its magnetic compensation temperature $T_{\rm comp}$, at which the reorientation of the sublattice magnetizations of GdIG was observed as the change of the heat-current direction in the SPE. \par 
\begin{figure}[tb]
\begin{center}
\includegraphics{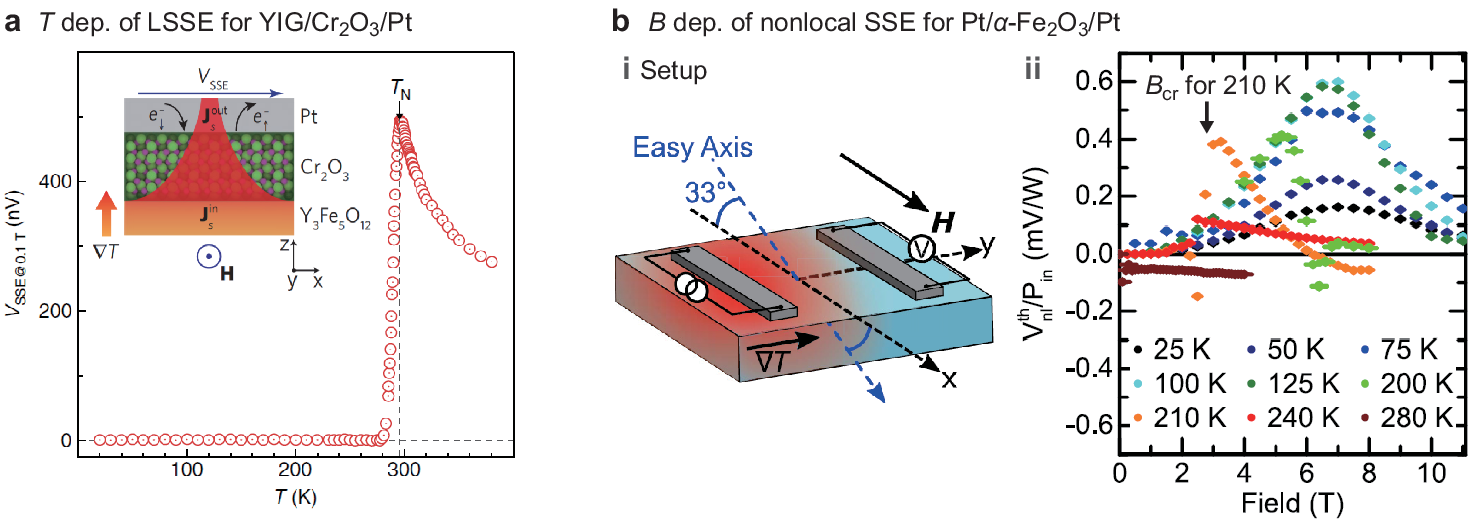}
\caption{
(a) $T$ dependence of the LSSE voltage $V_{\rm SSE}$ for an YIG/Cr$_2$O$_3$/Pt trilayer (see inset), where the $c$-axis of Cr$_2$O$_3$ is along the out-of-plane direction. With decreasing $T$, $V_{\rm ISHE}$ sharply drops to zero at $T_{\rm N}$ of Cr$_2$O$_3$.
Figure reproduced with permission from Reference \cite{Qiu2018NatMat}.  Copyright \copyright \, 2018 Springer Nature. 
(b) ({\it i}) Setup for a nonlocal SSE for a Pt/$\alpha$-Fe$_2$O$_3$/Pt sample. ({\it ii}) $T$ dependence of the nonlocal voltage normalized by the applied heating power $V^{\rm th}_{\rm nl}/P_{\rm in}$ at selected $T$. 
$B_{\rm cr}$ indicates a critical field at which ${\bf n}$ reorientates perpendicular to ${\bf B}$. 
Figure reproduced with permission from Reference \cite{Ross2021PRB}. Copyright \copyright \, 2021 American Physical Society. 
}\label{fig:Magnetic-ordering_SCMR}
\end{center}
\end{figure}
SSE-based detection of the N\'eel order and associated dynamics has also been reported through different approaches. 
In 2018, Qiu et al. \cite{Qiu2018NatMat} observed an on-off switching behavior in the LSSE voltage in YIG/Cr$_2$O$_3$/Pt trilayers across the N\'eel temperature $T_{\rm N} = 296~\textrm{K}$ ({\bf Figure \ref{fig:Magnetic-ordering_SCMR}a}).  
Here, below $T_{\rm N}$, the N\'eel vector ${\bf n}$ of Cr$_2$O$_3$ is pinned to the out-of-plane direction (easy-axis $||~c$) that is orthogonal to the spin component carried by magnons in the in-plane magnetized YIG film.
This prohibits magnon transmission from YIG to Cr$_2$O$_3$, and thus sharply suppresses the LSSE voltage below $T_{\rm N}$, being responsible for the observation \cite{Qiu2018NatMat}. 
In 2021, Ross et al. \cite{Ross2021PRB} demonstrated the nonlocal SSE mediated by the N\'eel vector ${\bf n}$ in $\alpha$-Fe$_2$O$_3$ under the external $B$ parallel to the easy-axis and the attached metal (Pt) strips ({\bf Figure \ref{fig:Magnetic-ordering_SCMR}b}, {\it subpanel i}), which allows for excluding significant contributions from the net magnetization vector \cite{Ross2021PRB}. In the easy-axis phase of $\alpha$-Fe$_2$O$_3$, close to but below the Morin transition temperature ($T_{\rm M} = 240~\textrm{K}$) such as 210 K, the nonlocal SSE voltage is initially constant around zero at a low $B$, but shows a sharp negative dip to positive peak transition across a critical field, $B_{\rm cr}$, where ${\bf n}$ reorientates perpendicular to ${\bf B}$ ({\bf Figure \ref{fig:Magnetic-ordering_SCMR}b}, {\it subpanel ii}). The result may be attributed to fluctuations of the N\'eel order as the magnetic anisotropy is compensated by the applied $B$ \cite{Ross2021PRB}. This work highlights the importance of the N\'eel-order magnon transport in the nonlocal SSE. 
We note that AF SSE induced by N\'eel dynamics is also demonstrated in the longitudinal configuration in $\alpha$-Fe$_2$O$_3$ and Cr$_2$O$_3$ \cite{Yuan2020APL,Li2020Nature,Reitz2020PRB}, which we discuss in Section \ref{sec:magnon-polarization}. 
In 2021, Luo and Liu et al. \cite{Luo2021PRB,Liu2021SciAdv} reported the LSSE-based detection of interfacial AF spin sublattices in Cr$_2$O$_3$ and their control by the applied electric field in the spin-flop phase at high magnetic fields, providing a new approach for controlling spin currents in AFMs. 
Very recently, Parsonnet et al. \cite{Parsonnet2022arXiv} reported non-volatile electric field control of the nonlocal SSE in multiferroic BiFeO$_3$ without external $B$ via the deterministic control of ferroelectric and magnetic order in BiFeO$_3$. \par 
%
%
\section{SSE AS PROBE OF SPIN CORRELATION}\label{sec:correlation}
As described in Section \ref{sec:theory}, in SSE the interfacial spin current is generated by the spin pumping, which indicates that SSE is sensitive to the transverse dynamical susceptibility or spin correlation function \cite{Xiao2010PRB,Barker2016PRL}. 
To test the concept, a paramagnetic phase just above the magnetic ordering temperature is an intriguing platform, at which the conventional magnon picture is no longer applicable, but short-range correlation of spin fluctuations may exist. \par
\begin{figure}[tb]
\begin{center}
\includegraphics{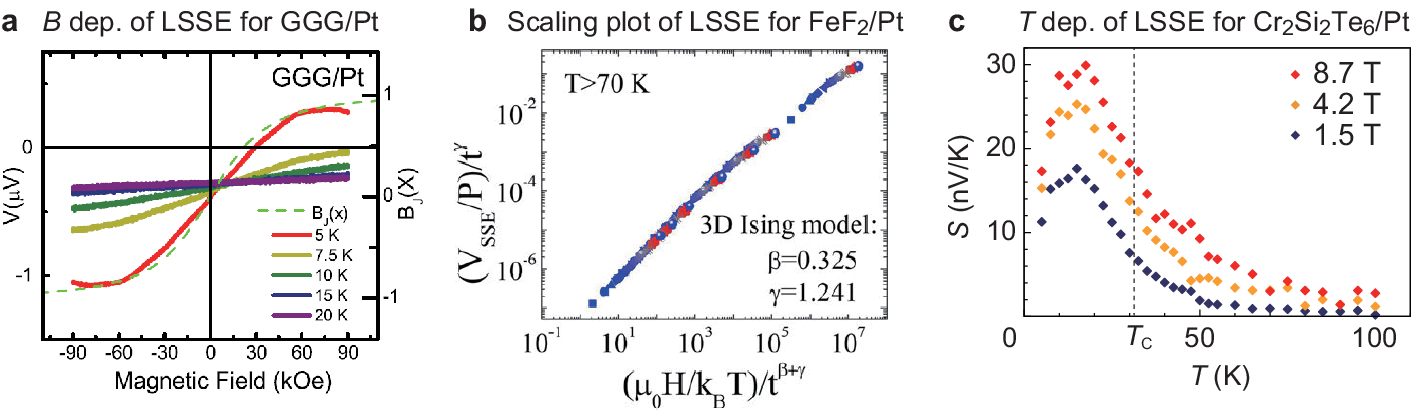}
\caption{
(a) $B$ dependence of the LSSE voltage $V$ for a Gd$_3$Ga$_5$O$_{12}$(GGG)/Pt system at selected $T$ and the Brillouin function $B_J$ for $S=7/2$ at 5 K. 
Figure reproduced with permission from Reference \cite{Wu2015PRL}. Copyright \copyright \, 2015 American Physical Society. 
(b) Scaling plot of the LSSE for a FeF$_2$/Pt system above $T_{\rm} = 70~\textrm{K}$. $V_{\rm SSE}$, $P$, and $t$ represent, respectively, the SSE voltage, applied heating power, and the reduced temperature: $t = (T-T_{\rm N})/T_{\rm N}$ for $T > T_{\rm N}$. 
Figure reproduced with permission from Reference \cite{Li2019PRL}. Copyright \copyright \, 2019 American Physical Society. 
(c) $T$ dependence of the LSSE voltage $S=E_{\rm ISHE}/\nabla T$ for a Cr$_2$Si$_2$Te$_6$/Pt system at selected $B$.
Figure reproduced with permission from Reference \cite{Ito2019PRB}. Copyright \copyright \, 2019 American Physical Society. 
}\label{fig:Spin-Correlation}
\end{center}
\end{figure}

Wu et al. \cite{Wu2015PRL,Liu2018PRB} investigated the LSSEs in the paramagnetic phase of Gd$_3$Ga$_5$O$_{12}$ (GGG) and DyScO$_3$ ({\bf Figure \ref{fig:Spin-Correlation}a}).
GGG does not exhibit long-range magnetic ordering down to $T \ll |\Theta_{\rm CW}|$ ($\Theta_{\rm CW} = - 2.3~\textrm{K}$ is the Curie-Weiss temperature \cite{Wu2015PRL}), but shows signatures of short-range correlations up to at least 5 K. DyScO$_3$ is a rather conventional AFM with the low $T_{\rm N}$ of 3.1 K. 
The observed SSE coefficient for GGG follows a scaling close to $T^{-1}$ as expected from the Curie-Weiss law $\left(\chi = C/(T - \Theta_{\rm CW}) \right)$. 
Besides, anisotropic $B$ dependencies in the LSSE for GGG were observed, and discussed in terms of short-range magnetic order due to geometrical frustration of GGG \cite{Liu2018PRB}. 
As a side note, owing to its large Gd$^{3+}$ spin of 7/2, GGG may show strong dipolar interaction under high $B$, generating collective spin-wave excitations, which may support bulk spin transport in GGG \cite{Oyanagi2019NatCommun}. \par
In 2019, Li et al. \cite{Li2019PRL} reported careful experiments and analysis on the AF LSSE in FeF$_2$ around its phase transition temperature ($T_{\rm N} = 70~\textrm{K}$). 
The experimental SSE coefficient near and above $T_{\rm N}$ follows the
critical scaling law with the critical exponents for magnetic susceptibility of 3D Ising systems, rather than the field-induced sublattice magnetization ({\bf Figure \ref{fig:Spin-Correlation}b}). 
This work clearly demonstrates that SSE is capable of probing correlations of spin fluctuations in magnetic systems \cite{Li2019PRL}. In the same year, Yamamoto et al. \cite{Yamamoto2019PRB} developed a linear response theory of AF SSE at elevated temperatures, predicting a cusp structure at $T_{\rm N}$ consistent with the experiment by Li et al. \cite{Li2019PRL}. \par 
It is worth mentioning that an insertion of an AF (NiO, CoO) film between FI (YIG) and NM (Pt, Ta, etc.) increases the LSSE signal especially at around $T_{\rm N}$, which is attributed to the enhanced spin conductance in the AF spacer around its $T_{\rm N}$ \cite{Lin2016PRL,Chen2016PRB,Prakash2016PRB}. This feature is also seen in an YIG/Cr$_2$O$_3$/Pt system at $T_{\rm N}=296~\textrm{K}$ of Cr$_2$O$_3$ ({\bf Figure \ref{fig:Magnetic-ordering_SCMR}a} \cite{Qiu2018NatMat}). \par 
Van der Waals 2D materials provide a fertile ground to study the effect of anisotropic spin correlation on SSE. 
Ito et al. \cite{Ito2019PRB} measured LSSEs in 2D layered FIs Cr$_2$Si$_2$Te$_6$ ($T_{\rm C} \sim 31~\textrm{K}$) and Cr$_2$Ge$_2$Te$_6$ ($T_{\rm C} \sim 65~\textrm{K}$) that exhibit in-plane short-range ferromagnetic correlations up to at least 300 K (200 K) for Cr$_2$Si$_2$Te$_6$ (Cr$_2$Ge$_2$Te$_6$), whereas out-of-plane correlations disappear slightly above $T_{\rm C}$ \cite{Williams2015PRB,Spachmann2022arXiv}. 
The LSSEs turned out to persist above $T_{\rm C}$, but disappear around 50 K (90 K) for a Cr$_2$Si$_2$Te$_6$/Pt (Cr$_2$Ge$_2$Te$_6$/Pt) system ({\bf Figure \ref{fig:Spin-Correlation}c}).  
The results show that the in-plane ($\perp \nabla T$) spin correlations do not solely produce the LSSE voltage; spin transport ($||~\nabla T$) between the planes enabled by the out-of-plane correlations is important to create nonequilibrium magnon population that leads to a finite interfacial spin current \cite{Ito2019PRB}. \par  
\begin{figure}[tbh]
\begin{center}
\includegraphics{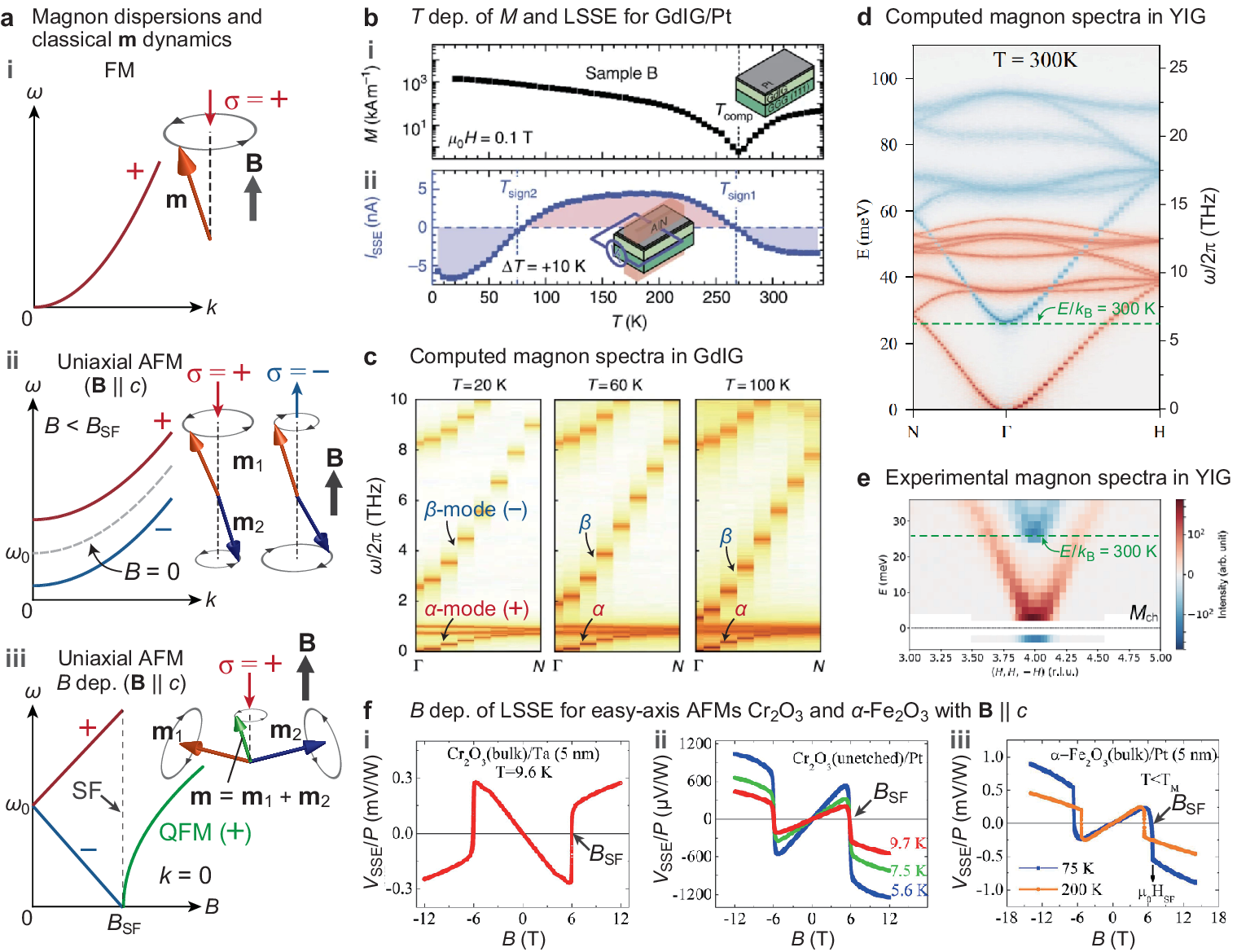}
\caption{
(a) Magnon dispersion relations and corresponding magnetization ${\bf m}$ dynamics for ({\it i}) a ferromagnet (FM) and ({\it ii}) a uniaxial antiferromagnet (AFM) under ${\bf B}~||~c$ (easy-axis) below the spin-flop (SF) field $B_{\rm SF}$.  
({\it iii}) Magnon gap at $k=0$ versus $B$ for ${\bf B}~||~c$. For $B>B_{\rm SF}$, the quasi-ferromagnetic (QFM) mode appears \cite{Yuan2020APL,Li2020Nature}. 
$\sigma = +$ $(-)$ represents the positive (negative) magnon polarization. 
(b) $T$ dependence of ({\it i}) $M$ in a GdIG film and  ({\it ii}) the LSSE signal $I_{\rm SSE} = V_{\rm ISHE}/R_{\rm Pt}$ ($R_{\rm Pt}$: Pt resistance).
(c) Magnon spectra of GdIG computed with ASD modelling. 
(b),(c) Figure reproduced with permission from Reference \cite{Geprags2016NatCommun}. Copyright \copyright \, 2016 Authors, licensed under a Creative Commons Attribution (CC BY) license. 
(d),(e) Magnon spectra of YIG at room temperature (d) obtained by implementing a quantum thermostat into ASD modelling and (e) measured with inelastic polarized neutron scattering. 
Dashed green lines show the thermal energy $k_{\rm B}T$ at 300 K. 
(d) Figure reproduced with permission from Reference \cite{Barker2019PRB}. Copyright \copyright \, 2019 American Physical Society. 
(e) Figure reproduced with permission from Reference \cite{Nambu2020PRL}. Copyright \copyright \, 2020 American Physical Society. 
(f) $B$ (${\bf B}~||~c$) dependence of the LSSE voltage normalized by the heating power $V_{\rm SSE}/P$ for easy-axis AFMs ({\it i}, {\it ii}) Cr$_2$O$_3$ and ({\it iii}) $\alpha$-Fe$_2$O$_3$ below $T_{\rm M}$ with metal (Ta, Pt) contacts. Figure reproduced with permission from Reference \cite{Yuan2020APL}. Copyright \copyright \, 2020 AIP Publishing.
} \label{fig:Magnon-polarization}
\end{center}
\end{figure}
\section{SSE AS PROBE OF MAGNON POLARIZATION}\label{sec:magnon-polarization}
A magnon may possess a positive ($+$) or negative ($-$) polarization, depending on whether the magnon mode carries a $+\hbar$- or $-\hbar$-spin angular momentum. In a classical picture, the  $+$ ($-$) polarization corresponds to the counterclockwise (clockwise) rotation of a magnetic moment ${\bf m}$ around the applied field ${\bf B}$ \cite{Nambu2020PRL}.  
In a ferromagnet, magnons have a single $+$ polarization ({\bf Figure \ref{fig:Magnon-polarization}a}, {\it subpanel i}), whereas in a uniaxial AFM there are two magnon modes with $+$ and $-$ polarizations that are degenerated at zero field ({\bf Figure \ref{fig:Magnon-polarization}a}, {\it subpanel ii}). 
In a ferrimagnet, both $+$ and $-$ polarized magnon modes exist due to the opposite sublattice moments that are split by the exchange field between the sublattices, and the $-$ polarized mode has higher energy than the $+$ one \cite{Barker2021JPSJ}. \par 
In 2016, Gepr\"ags et al. \cite{Geprags2016NatCommun} showed that SSE can be a measure of the magnon polarization. 
They observed two sign changes in the LSSE voltage for a GdIG/Pt system at $T_{\rm sign1} \sim 270~\textrm{K}$ and  $T_{\rm sign2} \sim 75~\textrm{K}$ ({\bf Figure \ref{fig:Magnon-polarization}b}). 
The higher one at $T_{\rm sign1}$ corresponds to the compensation point $T_{\rm comp}$ of GdIG ({\bf Figure \ref{fig:Magnon-polarization}b}, {\it subpanel i}), at which the Gd$^{3+}$ and Fe$^{3+}$ sublattice moments reverse in the presence of $B$, so does the SSE signal \cite{Geprags2016NatCommun}. 
The sign reversal at $T_{\rm sign2}$, by contrast, does not correspond to any changes in magnetic order. 
Theoretical modelling based on atomistic spin dynamics (ASD) shows that the sign change at $T_{\rm sign2}$ can be explained in terms of the $T$ dependent magnon modes with opposite polarization ($\alpha$- and $\beta$-modes in {\bf Figure \ref{fig:Magnon-polarization}c}) and their thermal occupation \cite{Geprags2016NatCommun}. 
At the low $T < T_{\rm sign2}$, the positively-polarized almost gapless $\alpha$-mode governs the sign of the LSSE in GdIG. As $T$ increases, the excitation gap for the negatively-polarized $\beta$-mode decreases, and the mode is thermally occupied (see {\bf Figure \ref{fig:Magnon-polarization}c}). For  $T > T_{\rm sign2}$, the spin-current contribution from the $\beta$-mode may be greater than that from the $\alpha$-mode, causing the sign reversal at $T_{\rm sign2}$ \cite{Geprags2016NatCommun}. \par
In most theories and experiments, YIG has been modeled as a ferromagnet with a single parabolic magnon mode \cite{Barker2016PRL}. 
However, due to its large magnetic primitive cell with localized 20 Fe$^{3+}$ moments, complicated 20 magnon modes exist, in fact. 
The spectra were recently computed with finite-temperature ASD modelling \cite{Barker2016PRL,Barker2019PRB} and measured with inelastic polarized neutron scattering \cite{Nambu2020PRL} ({\bf Figure \ref{fig:Magnon-polarization}d}, {\bf e}).
Interestingly, the gap of an optical mode with the negative polarization is comparable to the thermal energy $k_{\rm B}T$ at 300 K, which can be occupied at and above room temperature (see the green dashed lines representing $k_{\rm B}T$ at 300 K in {\bf Figure \ref{fig:Magnon-polarization}d}, {\bf e}). 
Calculations show that the optical mode reduces the spin pumping and SSE in YIG at room temperature and beyond due to its opposite polarization to the ferromagnetic acoustic mode \cite{Barker2016PRL,Nambu2020PRL}, which may cause a rapid decrease of SSE in YIG above room temperature, faster than the magnetization \cite{Uchida2014PRX}. \par 
Easy-axis AFMs are also a playground to investigate the impact of the magnon polarization on SSEs. 
In 2020, Li et al. \cite{Yuan2020APL,Li2020Nature} reported AF LSSEs in Cr$_2$O$_3$ and $\alpha$-Fe$_2$O$_3$ (in its easy-axis phase below $T_{\rm M}$) and found that the sign of the signals below the spin-flop field $B_{\rm SF}$ is opposite to that of ferromagnetic SSEs ({\bf Figure \ref{fig:Magnon-polarization}f}) \cite{Schreier2015JPhysD}. The result can be interpreted as the spin current from the negatively-polarized AF magnon mode (blue solid lines in {\bf Figure \ref{fig:Magnon-polarization}a}, {\it subpanels ii}, {\it iii}), whose excitation gap $\omega_0$ decreases with $B$. In the SF phase $B>B_{\rm SF}$, the signal changes sign, at which the SSE is dominated by the quasiferromagnetic mode having the positive polarization (see {\bf Figure \ref{fig:Magnon-polarization}f}, {\bf a}, {\it subpanel iii}). They also found that device surface treatment affects the sign reversal behavior \cite{Yuan2020APL,Li2020Nature}, which may explain the previously observed AF LSSEs showing the same sign as the ferromagnetic case \cite{Seki2015PRL,Wu2016PRL}. \par 
\section{SSE AS PROBE OF MAGNON-PHONON HYBRIDIZATION AND SCATTERING RATE} \label{sec:mpSSE}
\begin{figure}[tb]
\begin{center}
\includegraphics{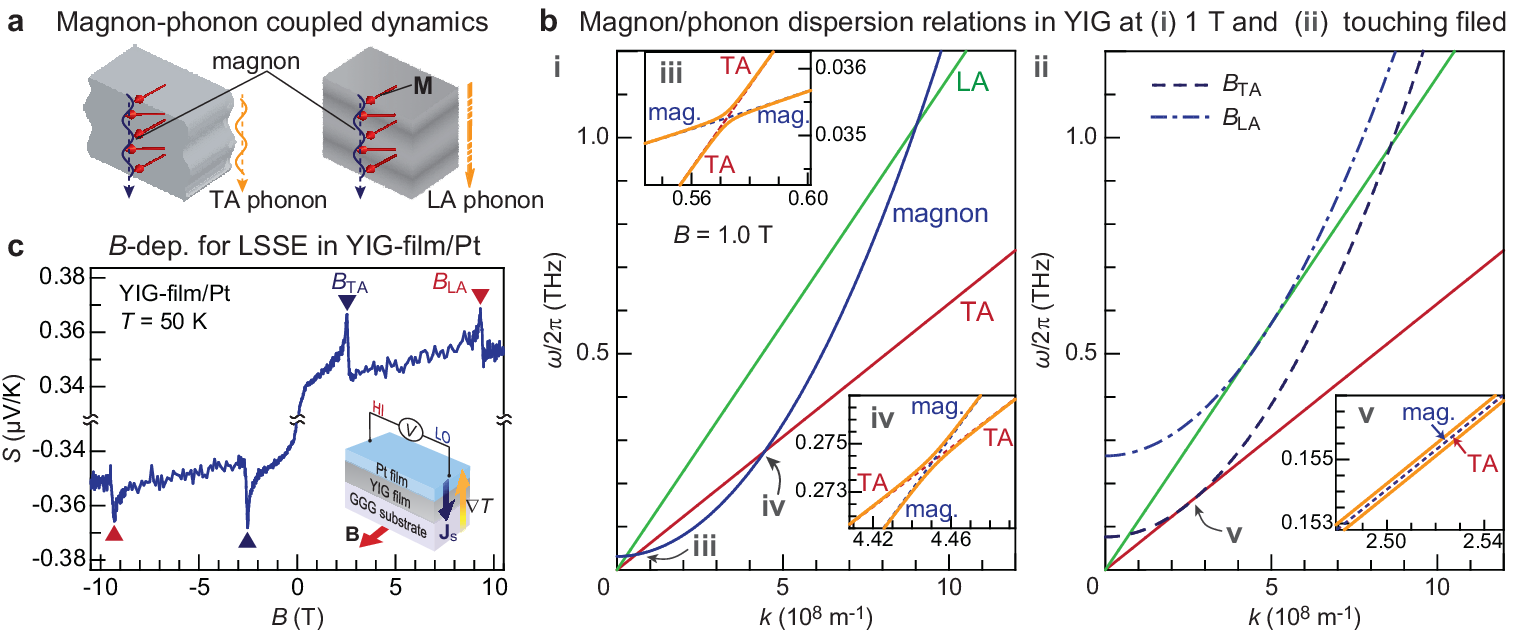}
\caption{(a) Magnon-phonon coupled dynamics. 
(b) Acoustic magnon, TA-, and LA-phonon dispersion relations in YIG at ({\it i}) $B=1.0~\textrm{T}$ and ({\it ii}) the touching field $B=B_{\rm TA}$ ($\sim 2.6~\textrm{T}$) and $B_{\rm LA}$ ($\sim 9.3~\textrm{T}$). 
({\it iii})-({\it v}) Blowups of the hybridized magnon-phonon (TA) modes at ({\it iii}, {\it iv})  $B=1.0~\textrm{T}$ and ({\it v}) $B=B_{\rm TA}$. 
(c) $B$ dependence of the LSSE voltage $S=E_{\rm ISHE}/\nabla T$ 
for an YIG-film/Pt system (see inset) at $T=50~\textrm{K}$.  
(a),(c) Figure reproduced with permission from Reference \cite{Kikkawa2016PRL}. Copyright \copyright \, 2016 American Physical Society. 
} \label{fig:MP-SSE}
\end{center}
\end{figure}

Recent experiments have revealed that hybridized magnon-phonon excitations induced by magnetoelastic coupling, i.e., magnon polarons ({\bf Figure \ref{fig:MP-SSE}a}, {\bf b}, {\it subpanels iii}, {\it iv}, {\it v}), are detected via $B$-dependent SSE voltages. Interestingly, the magnon-polaron signals contain unique information about relative scattering strengths of magnons and phonons  \cite{Kikkawa2016PRL,Flebus2017PRB}. \par
\begin{figure}[tbh]
\begin{center}
\includegraphics[width=16cm]{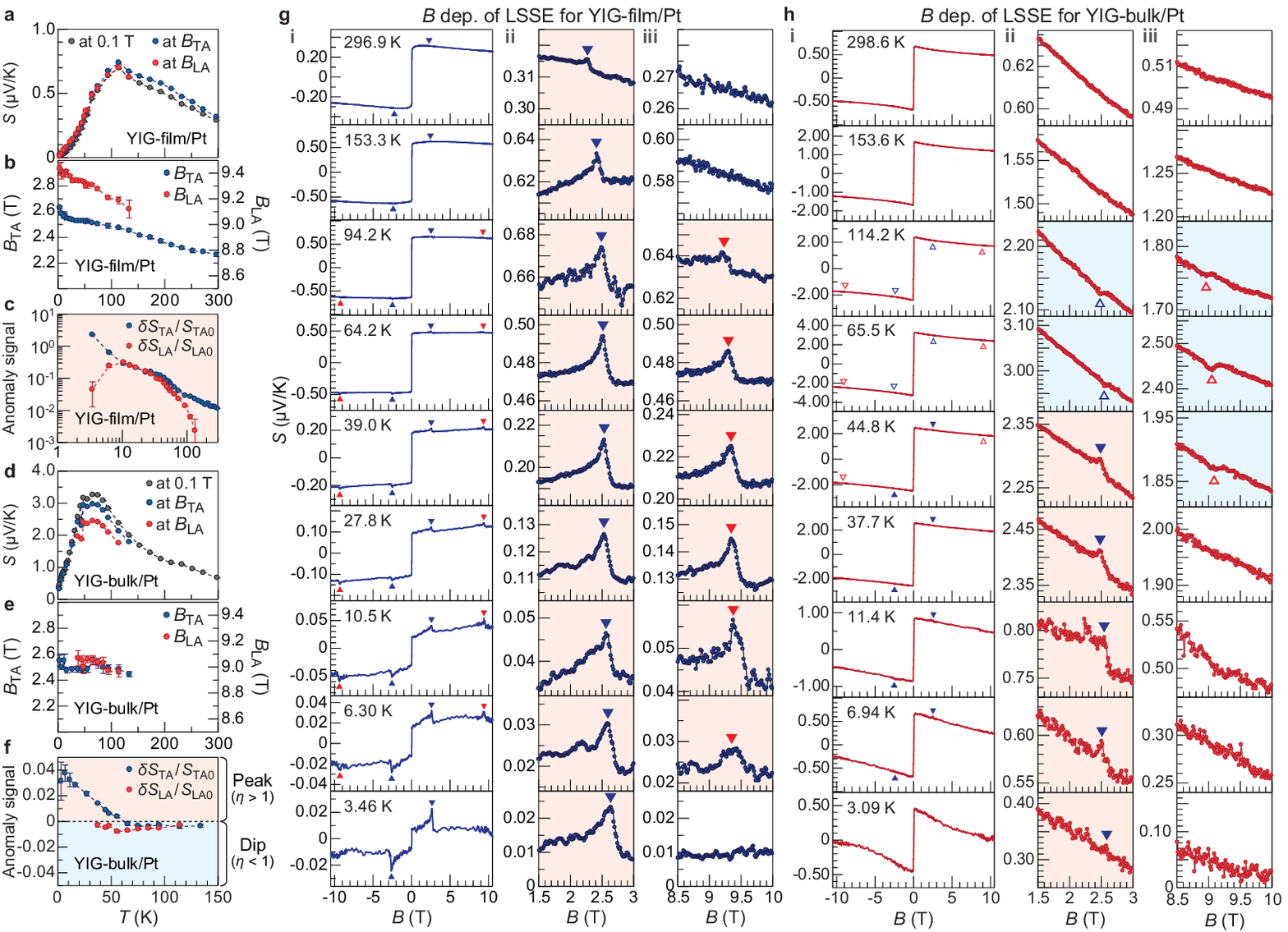}
\caption{(a),(b),(c) $T$ dependence of the (a) LSSE voltage $S=E_{\rm ISHE}/\nabla T$, (b) anomaly fields $B_{\rm TA}$ and $B_{\rm LA}$, and (c)  anomaly signal relative to the background one $\delta S_i/S_{i0} = \left(S(B_i) - S_{i0} \right)/S_{i0}$ for an YIG-{\it film}/Pt system.
Here, $S(B_i)$ and $S_{i0}$ represent, respectively, the $S$ and extrapolated background $S$ intensity at $B_i$ ($i = $ TA or LA) (see Reference \cite{Ramos2019NatCommun} for details of the definition of $\delta S_i/S_{i0}$).  
(d),(e),(f) $T$ dependence of (d) $S$, (e) $B_{\rm TA}$ and $B_{\rm LA}$, and (f) $\delta S_i/S_{i0}$ ($i = $ TA or LA) for an YIG-{\it bulk}/Pt system. 
(g),(h) $B$ dependence of (g,{\it i},h,{\it i}) $S$ for the (g,{\it i}) YIG-{\it film}/Pt and (h,{\it i}) YIG-{\it bulk}/Pt at selected $T$ and blowups of $S$ around (g,{\it ii},h,{\it ii}) $B_{\rm TA}$ and (g,{\it iii},h,{\it iii}) $B_{\rm LA}$.
$S$ reduction at high $B$ originates from the freeze-out of magnons by the Zeeman effect \cite{Kikkawa2015PRB,Jin2015PRB,Kikkawa2016PRL}. 
The red (blue) shadowed areas show the regions where magnon-polaron peak (dip) signals are observed, indicating $\eta > 1$ ($\eta < 1$).
Data in (a) and (g) below $T=50~\textrm{K}$ are reproduced with permission from Reference \cite{Kikkawa2016PRL}. Copyright \copyright \, 2016 American Physical Society. 
} \label{fig:MP-SSE_film-vs-bulk}
\end{center}
\end{figure}

A high-resolution field scan discerns saw-tooth peak structures in the LSSE for an YIG-film/Pt bilayer at $B_{\rm TA} \sim 2.6~\textrm{T}$ and $B_{\rm LA} \sim 9.3~\textrm{T}$ as marked by the blue and red triangles in {\bf Figure \ref{fig:MP-SSE}c}, respectively \cite{Kikkawa2016PRL}. 
The SSE anomalies show up when the acoustic magnon mode in YIG shifts upward with external $B$ such that TA and LA phonon dispersions become tangential at $B_{\rm TA}$ and $B_{\rm LA}$ ({\bf Figure \ref{fig:MP-SSE}b}). 
Under these ``touching'' conditions, the magnon and phonon modes can be coupled over the largest volume in momentum space, leading to the maximal effect of magnon-polaron formation in magnonic spin transport (compare the magnon-phonon (TA) hybridized regions for the ``intersect'' situation at $B = 1~\textrm{T}$ ({\bf Figure \ref{fig:MP-SSE}b}, {\it subpanels iii}, {\it iv}) and the ``touching'' one $B = B_{\rm TA}$ ({\bf Figure \ref{fig:MP-SSE}b}, {\it subpanel v})).  
The Boltzmann theory \cite{Flebus2017PRB} revealed that, when the scattering rate of magnons $\tau^{-1}_{\rm mag}$ is larger (smaller) than that of phonons $\tau^{-1}_{\rm ph}$, magnon polarons may have a longer (shorter) lifetime than pure magnons and thus enhance (suppress) the SSE at the touching fields. 
Clear peaks observed in a wide $T$ range for the YIG film ({\bf Figure \ref{fig:MP-SSE}c}, {\bf Figure \ref{fig:MP-SSE_film-vs-bulk}c}, {\bf g}) suggest higher acoustic than magnetic quality of the sample, i.e., $\eta = \tau^{-1}_{\rm mag}/\tau^{-1}_{\rm ph} > 1$ \cite{Kikkawa2016PRL,Flebus2017PRB}. \par
In general, scattering rates for magnons and phonons depend on microscopic scattering mechanism, temperature, wavenumber $k$, and sample qualities \cite{Flebus2017PRB,Streib2019PRB,Shi2021PRL}. 
Indeed, for an YIG-bulk/Pt system, the anomalies take both peak and dip shapes depending on $T$ and the touching fields $B_{\rm TA}$ and $B_{\rm LA}$ (see {\bf Figure \ref{fig:MP-SSE_film-vs-bulk}f}, {\bf h}), showing that $T$-, $k$-, and $B$-dependent magnon and phonon scattering mechanisms should be taken into consideration. Very recently, Shi et al. \cite{Shi2021PRL} reported detailed experimental and theoretical studies on magnon-polaron anomalies in YIG-bulk/Pt systems and showed that $T$-dependent 4-magnon scattering may lead to dip-to-peak transition (from $\eta > 1$ to $\eta < 1$). \par 
So far, magnon-polaron anomalies in SSEs have been reported in both the longitudinal and nonlocal configurations for various magnets such as ferrimagnetic YIG \cite{Kikkawa2016PRL,Cornelissen2017PRB,Oyanagi2020AIPAdv}, Fe$_{3}$O$_{4}$ \cite{Xing2020PRB}, NiFe$_{2}$O$_{4}$ \cite{Shan2018APL}, Ni$_{0.65}$Zn$_{0.35}$Al$_{0.8}$Fe$_{1.2}$O$_{4}$ \cite{Wang2018APL}, (partially) compensated ferrimagnetic Lu$_{2}$Bi$_{1}$Fe$_{4}$Ga$_{1}$O$_{12}$ \cite{Ramos2019NatCommun} and Gd$_3$Fe$_5$O$_{12}$ \cite{Yang2021PRB}, and AF Cr$_2$O$_3$ \cite{Li2020PRL_Cr2O3}. The observation of magnon-polaron anomalies in Cr$_2$O$_3$ below $B_{\rm SF}$ demonstrates that the SSE in the system is indeed governed by the negatively-polarized magnon mode, opposite to ferromagnetic magnons, giving insight to unraveling the origin of the sign of AF SSEs \cite{Reitz2020PRB,Yamamoto2022PRB}. \par
It is worthwhile to mention that sufficiently strong magnon-magnon and phonon-phonon scattering processes may destroy the magnon-phonon coherence.  
Schmidt et al. \cite{Schmidt2018PRB} formulated a Boltzmann transport theory in such a parameter regime and showed that similar SSE anomalies manifest through the ``phonon drag'' process at the touching fields.  
In 2021, Schmidt and Brouwer \cite{Schmidt2021PRB} showed a detailed calculation on the low-$T$ LSSE which treats exactly the frequency dependence of the magnon and phonon distribution functions under various scattering mechanisms, beyond the conventional approach based on a Planck-type or Bose-Einstein-type ansatz. 
They found that, in the low $T$ below $\sim 10~\textrm{K}$ and in sample sizes relevant to experiments, thermal spin transport in YIG may be dominated by magnon-polaron modes, which sharply enhances the SSE at the touching field \cite{Schmidt2021PRB}. \par 
%
%
\section{SPIN SEEBECK EFFECTS IN QUANTUM MAGNETS} \label{sec:qSSE}
%
SSE has been studied not only in conventional ferro-, ferri-, and antiferro-magnets, but also in exotic materials called quantum spin liquids (QSLs) with strong quantum fluctuations. Here, QSL in a broader sense is a state of a magnet in which spin correlations are present, while long-range magnetic ordering is absent due to quantum fluctuations reinforced via the low-dimensionality and frustration \cite{Hirobe2018JAP}. 
For such systems, collective excitations of localized spins are no longer conventional spin waves, or magnons, and more exotic spin excitations show up. 
Recent experiments have revealed that SSE serves as a powerful probe for spin correlation and transport in QSLs, including Tomonaga-Luttinger (TL) spin liquids \cite{Hirobe2017NatPhys,Hirobe2018JAP}, spin-nematic systems \cite{Hirobe2019PRL}, and spin-dimer systems \cite{Chen2021NatCommun_CuGeO3,Xing2022APL}. \par 
\subsection{Spinons in TL spin liquids} \label{sec:spinon} 
\begin{figure}[tbh]
\begin{center}
\includegraphics{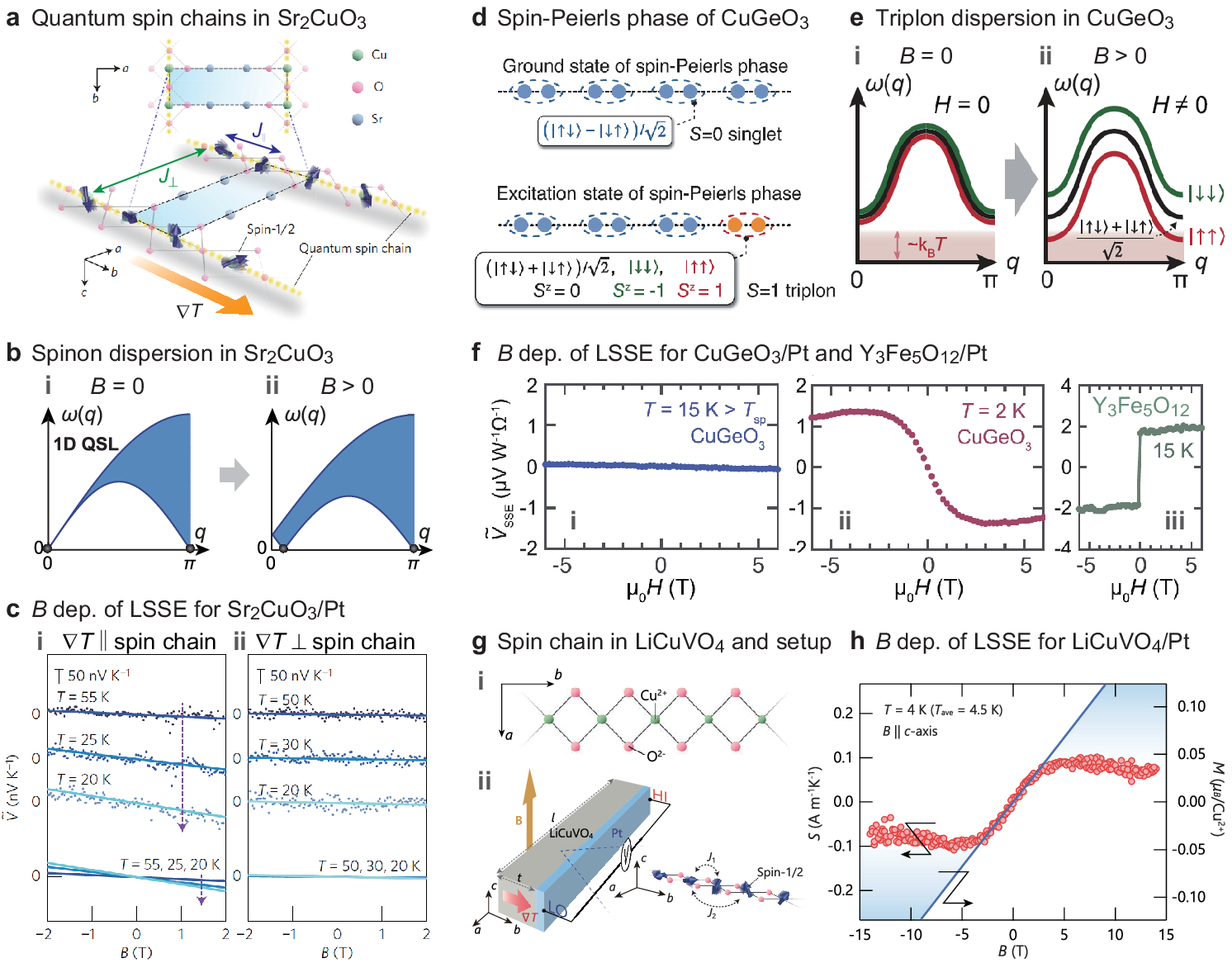}
\end{center}
\caption{(a) Quantum spin chains in Sr$_2$CuO$_3$ \cite{Hirobe2017NatPhys}.   
(b) Excitation spectrum, $\omega (q)$, of 1D spinons in Sr$_2$CuO$_3$ under ({\it i}) zero and ({\it ii}) finite $B$ \cite{Hirobe2018JAP}. Figure reproduced with permission from Reference \cite{Hirobe2018JAP}. Copyright \copyright \, 2018 AIP Publishing. 
(c) $B$ dependence of the LSSE voltage ${\tilde V}=V/\Delta T$ 
for Sr$_2$CuO$_3$/Pt systems at selected $T$ for ({\it i}) ${\nabla T}~||~$ spin chain and  ({\it ii}) ${\nabla T} \perp $ spin chain \cite{Hirobe2017NatPhys}. 
(a),(c) Figure reproduced with permission from Reference \cite{Hirobe2017NatPhys}. Copyright \copyright \, 2016 Springer Nature. 
(d) Ground and excited states of the spin-Peierls (SP) phase of CuGeO$_3$. 
(e) $\omega (q)$ of triplons in CuGeO$_3$ under ({\it i}) zero and ({\it ii}) finite $B$ \cite{Chen2021NatCommun_CuGeO3}. 
(f) $B$ dependence of the LSSE voltage ${\tilde V}_{\rm SSE}$ (the voltage $V$ normalized by the applied heating power $P$ and Pt resistance $R_{\rm Pt}$) for a CuGeO$_3$/Pt system at ({\it i}) $T=15~\textrm{K}$ and ({\it ii}) $2~\textrm{K}$ and ({\it iii}) for a Y$_3$Fe$_5$O$_{12}$/Pt  system at $15~\textrm{K}$ \cite{Chen2021NatCommun_CuGeO3}. 
(d)-(f) Figure reproduced with permission from Reference \cite{Chen2021NatCommun_CuGeO3}. Copyright \copyright \, 2021 Authors, licensed under a Creative Commons Attribution (CC BY) license. 
(g) ({\it i}) Spin chain in LiCuVO$_4$ and ({\it ii}) Setup for detecting the LSSE in a LiCuVO$_4$/Pt system. 
(h) $B$ dependence of the LSSE voltage ${\tilde S}=E_{\rm ISHE}/(\rho_{\rm Pt} \nabla T)$ ($\rho_{\rm Pt}$: Pt resistivity) 
for a LiCuVO$_4$/Pt system and $M$ in LiCuVO$_4$ at $T=4~\textrm{K}$. 
(g),(h) Figure reproduced with permission from Reference \cite{Hirobe2019PRL}. Copyright \copyright \, 2019 American Physical Society. 
} \label{fig:SSEs_quantum_spins}
\end{figure}

Spinons generally refer to magnetic elementary excitations in QSLs \cite{Hirobe2017NatPhys}. 
The most typical example is found in 1D spin-1/2 chains realized in some oxides such as a Mott insulator Sr$_2$CuO$_3$ having 1D Cu$^{2+}$ spin ($S=1/2$) with large nearest-neighbor AF exchange coupling $J_{||}$ ($\sim 2000~\textrm{K}$, much stronger than the inter-chain coupling $J_{\perp}$; {\bf Figure \ref{fig:SSEs_quantum_spins}a}) \cite{Hirobe2017NatPhys}. 
Due to the 1D nature, its spin fluctuation is so strong, leading to a paramagnetic state with strong spin-singlet correlations. 
The spin excitation from this correlated ground state is particle-like, called a  ``spinon'', and has a gapless dispersion robust to external fields and magnetic anisotropies ({\bf Figure \ref{fig:SSEs_quantum_spins}b}). \par
In 2017, Hirobe et al. \cite{Hirobe2017NatPhys} demonstrated the spinon SSE in a 1D TL spin liquid state in Sr$_2$CuO$_3$ with Pt contact. 
The $T$ and crystallographic orientation dependencies indicate that the observed LSSE originates from the 1D spin correlation along the chain ({\bf Figure \ref{fig:SSEs_quantum_spins}c}). The polarity of the signal is opposite to that due to ferromagnetic SSEs, showing that classical spin fluctuations are not responsible for the experimental results.   
The results were reproduced by adopting the Bethe ansatz into a general formulation of SSE \cite{Hirobe2017NatPhys}, showing that the signal polarity is determined by finite-temperature dynamics of 1D spinons. \par 
\subsection{Triplons in dimerized spin systems}\label{sec:triplon} 
Among quantum spin systems without magnetic order, dimerized spin systems are an important playground, in which two neighboring spins with AF coupling are frozen as $S = 0$ singlets in the ground state $[(\mid \uparrow\downarrow \rangle-\mid \downarrow\uparrow \rangle)/\sqrt{2}]$. The elementary spin excitations are $S = 1$ triplet states $[ \mid \downarrow\downarrow \rangle$, $(\mid \uparrow\downarrow \rangle+\mid \downarrow\uparrow \rangle)/\sqrt{2}$, and $\mid \uparrow\uparrow \rangle ]$, called a ``triplon'' ({\bf Figure \ref{fig:SSEs_quantum_spins}d}) \cite{Chen2021NatCommun_CuGeO3}. \par   
Recently, Chen et al. \cite{Chen2021NatCommun_CuGeO3} demonstrated the triplon SSE in a spin-Peierls (SP) material CuGeO$_3$ having a 1D Cu$^{2+}$ spin-1/2 chains with AF exchange interaction for nearest-neighbor spins. Below its SP transition temperature $T_{\rm SP} \sim 14.5~\textrm{K}$, the chain distorts so that the distance between neighboring spins alternates. The bond-alternating exchange interaction causes neighboring spins to dimerize to reduce the total energy, creating a gap in the excitation spectrum ($\omega_0 \sim 23~\textrm{K}$) (see the triplon dispersions sketched in {\bf Figure \ref{fig:SSEs_quantum_spins}e}) \cite{Chen2021NatCommun_CuGeO3}.  
The triplon states are threefold degenerated at $B=0$, whereas they are lifted to split into three different energy levels under $B\neq 0$, the lowest mode of which is the $\mid \uparrow\uparrow \rangle$ state carrying a spin polarization opposite to ferromagnetic magnons (same sign as spinons) \cite{Hirobe2017NatPhys,Hirobe2018JAP}. Chen et al. \cite{Chen2021NatCommun_CuGeO3} successfully observed the triplon LSSE in CuGeO$_3$/Pt systems in the SP phase, whose sign is opposite to that for YIG/Pt as expected ({\bf Figure \ref{fig:SSEs_quantum_spins}f}).  
In 2022, Xing et al. \cite{Xing2022APL} reported the LSSE in a spin-gapped quantum magnet Pb$_2$V$_3$O$_9$ having a relatively low critical field $B_{\rm c}$ to form the Bose--Einstein condensation (BEC) states of triplons, at which the excitation gap for $\mid \uparrow\uparrow \rangle$ is lower than the energy of ground state \cite{Chen2021NatCommun_CuGeO3,Xing2022APL}. 
They \cite{Xing2022APL} found that the LSSE voltages in Pb$_2$V$_3$O$_9$/Pt  are maximal at $B_{\rm c}$, whose $T$ dependence follows the power law for the BEC phase boundary: $T \propto (B - B_{\rm c})^{2/3}$  \cite{Xing2022APL}. 
\subsection{Spin-nematic TL liquids}\label{sec:spin-nematic}
A spin-nematic ordered phase is a physical state with a spin quadrupolar order and without any spin-dipolar (magnetic) order \cite{Sato2020PRB_spin-nematic}.   
The state typically emerges in a 1D frustrated spin-1/2 chain with the ferromagnetic nearest neighboring exchange interaction $J_1 < 0$ and the AF next nearest neighboring one $J_2 > 0$ ({\bf Figure \ref{fig:SSEs_quantum_spins}g}) \cite{Hirobe2019PRL}. 
As $B$ is increased in this system, the spin-nematic TL liquid state appears in a wide $B$ range. Here,
not only single magnons but also magnon pairs (molecules of two magnons) are excited, whose energy gap is equivalent to the binding energy of magnon pairs and zero, respectively. \par
Hirobe et al. \cite{Hirobe2019PRL} investigated the LSSE in a spin-nematic magnet  LiCuVO$_4$ with quasi-1D Cu$^{2+}$ spin-1/2 chains and observed a strong $B$-induced signal reduction ({\bf Figure \ref{fig:SSEs_quantum_spins}h}). 
They attributed the result to the suppressed interfacial spin exchange due to the stabilization of magnon-pairs carrying spin-2 which cannot contribute to the interfacial spin injection in SSE governed by the spin-1 exchange between single magnons and conduction electrons \cite{Hirobe2019PRL} (see {\bf Figure \ref{fig:SSE_mechanism}b}). 
The selective probing feature of spin-1 magnetic excitations for SSEs is distinct from thermal conductivity measurements, as the latter simultaneously probes phonons as well as multiple spin-1 and spin-2 excitations.
This study shows that SSE may detect signatures of spin-nematic states and their transport properties \cite{Hirobe2019PRL}.  \par
\begin{figure}[tbh]
\begin{center}
\includegraphics{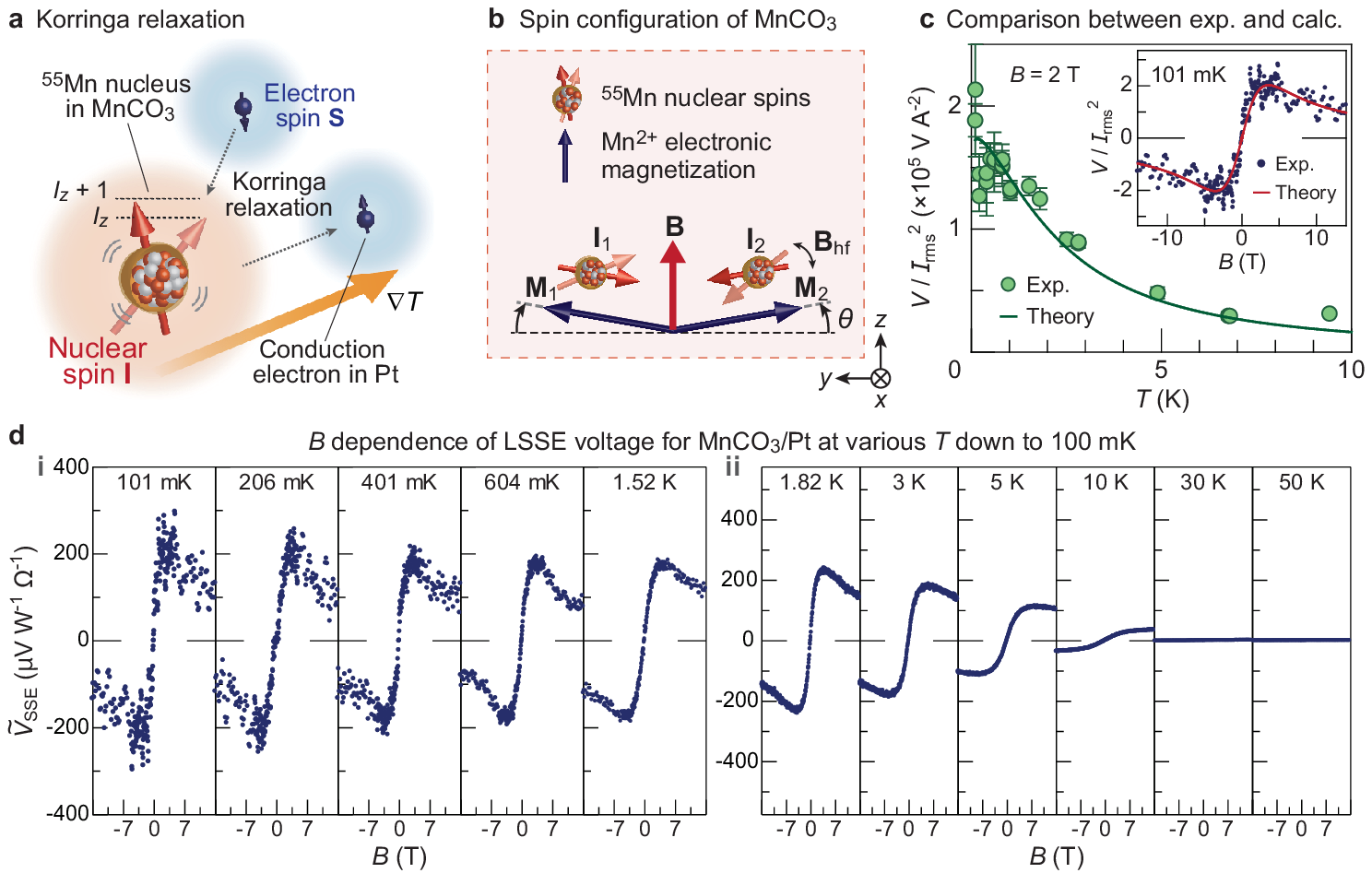}
\end{center}
\caption{(a) Nuclear SSE induced by the interfacial Korringa relaxation process, the spin-conserving flip-flop scattering between a nuclear spin, ${\bf I}$, of $^{55}$Mn in MnCO$_3$ and an electron spin, ${\bf S}$, in Pt via the interfacial hyperfine interaction. 
(b) Orientation of the canted Mn$^{2+}$ sublattice magnetizations ${\bf M}_1$ and ${\bf M}_2$ and the $^{55}$Mn nuclear spins ${\bf I}_1$ and ${\bf I}_2$ in MnCO$_3$ for ${\bf B}~||~(111)$.  
Due to the strong internal hyperfine field of $B_{\rm hf} \sim 60~\textrm{T}$, ${\bf I}_1$ and ${\bf I}_2$ orient (antiparallel) to the ${\bf M}_1$ and ${\bf M}_2$ directions, respectively.
(c) $T$ dependence of the LSSE voltage $V$ normalized by the applied current squared $I_{\rm rms}^2$ (proportional to the heating power $P$) for MnCO$_3$/Pt Device 1 at $B = 2~\textrm{T}$. The inset shows the $B$ dependence of $V/I_{\rm rms}^2$  at  $T = 101~\textrm{mK}$. Theoretical results for the nuclear SSE are also plotted with solid curves. 
(a)-(c) Figure reproduced from Reference \cite{Kikkawa2021NatCommun}. Copyright \copyright \, 2021 Authors, licensed under a Creative Commons Attribution (CC BY) license. 
(d) $B$ dependence of the LSSE voltage ${\tilde V}_{\rm SSE} =V_{\rm ISHE}/(PR_{\rm Pt})$ for ({\it i}) Device 1 for $100~\textrm{mK}<T<1.6~\textrm{K}$ and ({\it ii}) Device 2 for $1.8~\textrm{K}<T<50~\textrm{K}$.
} \label{fig:NSSE}
\end{figure}
%
\section{NUCLEAR-SPIN SEEBECK EFFECT} \label{sec:NSSE}
So far, we have discussed the SSEs caused by electron spins. At low temperatures or high magnetic fields, their efficiency inevitably disappears due to freeze-out of magnons \cite{Kikkawa2015PRB,Kikkawa2016JPSJ}.
In solids, however, there is an as-yet-unexplored entropy carrier thermally activated even in such an environment: a nuclear spin ({\bf Figure \ref{fig:NSSE}a}). 
The feature originates from the tiny gyromagnetic ratio of a nuclear spin $\gamma_{\rm n}$, $\sim 10^3$ times less than that of an electron $\gamma$, which makes its excitation gap in the range below 1 GHz ($\sim 50~\textrm{mK}$ in units of $T$) in ambient fields. 
Here a question arises: Is it possible that nuclear spins drive SSEs? 
Based on the ISHE measurements for MnCO$_3$/Pt systems at ultralow temperatures, we recently answered this question affirmatively \cite{Kikkawa2021NatCommun}. \par 
MnCO$_3$ is an easy-plane canted AF insulator having a large nuclear spin $I = 5/2$ of $^{55}$Mn nuclei and strong hyperfine coupling \cite{Shiomi2019NatPhys}. 
Below $T_{\rm N} = 35~\textrm{K}$, the Mn$^{2+}$ sublattice magnetizations ${\bf M}_1$ and ${\bf M}_2$ are aligned in the (111) plane and canted slightly from the collinear AF configuration due to the Dzyaloshinskii--Moriya interaction (see {\bf Figure \ref{fig:NSSE}b}). 
The hyperfine (Overhauser) fields $B_{\rm hf} $ acting on the $^{55}$Mn sublattice nuclear spins ${\bf I}_1$ and ${\bf I}_2$ from ${\bf M}_1$ and ${\bf M}_2$ reach as large as $\sim 60~\textrm{T}$ \cite{Shiomi2019NatPhys}, which reinforces the nuclear spin polarization ($\sim 40\%$ at 100 mK) and orients ${\bf I}_1$ and ${\bf I}_2$ along the ${\bf M}_1$ and ${\bf M}_2$ directions, respectively ({\bf Figure \ref{fig:NSSE}b}).  
Moreover, the canting angle $\theta$ increases with $B$, so does the net nuclear-spin polarization (${\bf I}_1 + {\bf I}_2$) along ${\bf B}$.
The advantage makes the nuclear SSE experimentally feasible.  
{\bf Figure \ref{fig:NSSE}d} shows the $B$ dependence of the LSSE voltage in MnCO$_3$/Pt systems for $100~\textrm{mK} < T < 50~\textrm{K}$. 
The observed SSE is enhanced down to 100 mK and is not suppressed under the strong field of 14 T (see also {\bf Figure \ref{fig:NSSE}c}). Importantly, even in this extreme environment, the nuclear-spin mode in MnCO$_3$ can be greatly excited because its excitation gap is as small as $\sim 30~\textrm{mK}$, which is little affected by the field. By contrast, electronic magnons freeze out by the Zeeman gap $\gamma B \sim 19~\textrm{K}$, much higher than the thermal energy. A nuclear SSE theory indeed quantitatively reproduces the experimental results, in which interfacial hyperfine coupling between nuclear spins in MnCO$_3$ and electrons in Pt is taken into account \cite{Kikkawa2021NatCommun} (see {\bf Figure \ref{fig:NSSE}a}, {\bf c}). \par 
\section{SUMMARY AND OUTLOOK} \label{sec:summary} 
In this article, we reviewed the recent progress of SSE research and discussed its emerging role as an instrument for magnon (phonon) excitations, transport, spin correlation, interfacial spin-exchange, static magnetic and N\'eel order, and domains. 
This unique feature is realized because both the interfacial spin-current injection and bulk spin transport play essential roles in SSEs, unlike other conventional spintronic phenomena that appear only in a nanoscale, highlighting the power of SSEs. \par  
There are some interesting theoretical proposals relevant to the present topics. 
Matsuo et al. \cite{Matsuo2018PRL} calculate the spin-current noise in spin pumping and SSE, which can be used to determine the effective spin carried by a magnon modified by interfacial spin-nonconserving processes and also to estimate $\theta_{\rm SH}$.
Nasu and Naka \cite{Nasu2021PRB} investigate SSEs in nonmagnetic excitonic insulators (NEIs) and conclude that SSE signals appear without external fields due to the time-reversal symmetry breaking inherent in the NEI state. 
Takikawa et al. \cite{Takikawa2022PRB} calculate a thermal spin current in a Kitaev spin liquid state and show that the SSE can be a measure of a chiral Majorana edge mode.  
Besides, magnon polarons discussed in this article may affect magnonic spin and thermal conductivities \cite{Flebus2017PRB} and also, in the nonlocal configuration, induce a Fulde-Ferrell-Larkin-Ovchinnikov (FFLO)-like oscillatory voltage as a function of injector-detector distance \cite{Rameshti2019PRB}.
All these theoretical works await experimental demonstration. \par
Machine learning is becoming a valuable tool to uncover hidden, complicated regularities in datasets. In 2019, Iwasaki et al. \cite{Iwasaki2019SciRep} demonstrated the utility of machine learning to elucidate the fundamental physics of SSE and optimize key material parameters to enhance the thermopower. Subsequently, they also developed interpretable machine learning \cite{Iwasaki2019npjComput Mater}, which indeed led to the discovery of a material with large spin-driven thermoelectric efficiency. A machine learning-based approach may therefore help not only in developing novel materials, but also in guiding theoretical studies in this field. \par 
Exploitation of SSEs in 2D van der Waals materials with tunable static and dynamical magnetic properties \cite{Gibertini2019NatNanotech,Mak2019NatRevPhys} is a fruitful avenue of investigation.
Some experiments have been reported recently aside from the LSSEs in Cr$_2$Si$_2$Te$_6$ and Cr$_2$Ge$_2$Te$_6$ \cite{Ito2019PRB} discussed in Section \ref{sec:correlation}. 
In 2019, Xing et al. \cite{Xing2019PRX} demonstrated long-distance magnon transport in a quasi-2D AF insulator MnPS$_3$ via a nonlocal SSE. 
Subsequently, the nonlocal SSE in this system is shown to be turned on and off by an electrical current though a metal gate due to the nonlinear gate dependence \cite{Chen2021NatCommun_MnPS3}.  
In 2020, Lee et al. \cite{Lee2020AFM_WSe2} showed that the insertion of a monolayer WSe$_2$ between YIG and Pt layers enhances the LSSE voltage by a factor of $\sim 5$ compared to that in an YIG/Pt system, showing a new avenue on SSE research with  2D transition dichalcogenide materials. 
In the same year, Liu et al. \cite{Liu2020PRB} demonstrated the nonlocal SSE in a 2D ferromagnetic CrBr$_3$ flake ($\sim$ 10 layers) fully encapsulated by two layers of h-BN, which paves the way for future magnonic devices \cite{Chumak2022IEEE} with air-sensitive 2D magnets.  
Studies on SSEs with 2D materials have just begun, and more fascinating phenomena are to be observed through control of the layer number and stacking combinations of 2D materials in the near future. \par  
%
%
\section*{DISCLOSURE STATEMENT} 
The authors are not aware of any affiliations, memberships, funding, or financial holdings that might be perceived as affecting the objectivity of this review. 
%
\section*{ACKNOWLEDGMENTS}
The authors thank many colleagues and collaborators for fruitful discussions.  
This work was supported by JST-CREST (JPMJCR20C1 and JPMJCR20T2),  
Grant-in-Aid for Scientific Research (JP19H05600 and JP20H02599) from JSPS KAKENHI, Japan, 
Institute for AI and Beyond of the University of Tokyo, 
IBM-UTokyo lab, 
and Daikin Industries, Ltd. 
\end{document}